\documentclass[draft,showpacs,preprintnumbers,amsmath,amssymb,fourier]{revtex4-2}
\usepackage{graphicx}
\usepackage{float}
\usepackage{color}

\usepackage{amsmath, amsthm, amsfonts}
\usepackage{amssymb}

\theoremstyle{definition}

\theoremstyle{remark}

\begin{document}
	
	\title{Scale-Dependent Weak-Field Dynamics in Horndeski Theory: Galaxy Rotation Curves and Turnaround Scales}

	\begin{abstract}
		In this work, the weak-field dynamics of a gravitational-wave compatible subclass of Horndeski gravity is investigated on galactic and intergalactic scales. The analysis is performed within the weak-field approximation for stationary and approximately spherically symmetric configurations, retaining the nonlinear scalar derivative interactions responsible for scale-dependent modifications of the gravitational field. On galactic scales, the transition regime is investigated through the effective potential, the stability of circular orbits, galaxy rotation curves, and the relation between dynamical and lensing masses. The analysis shows that the scalar field modifies the orbital dynamics and the radial dependence of the rotational velocity, while generating a nonvanishing gravitational slip between the dynamical and lensing mass estimates. On intergalactic scales, the asymptotic weak-field dynamics is investigated and a turnaround scale is derived from the balance between the effective gravitational attraction and the large-scale curvature contribution. At large distances, the asymptotic gravitational field is governed by the background curvature contribution. 
	\end{abstract}
	
	
	\author{Hatice Özer}
	\email[Corresponding author: ]{hatice.ozer@istanbul.edu.tr}
	
	\affiliation{Department of Physics, Faculty of Sciences, Istanbul University,
		34134 Istanbul, Türkiye\\
		ORCID: 0000-0002-6325-0598}
	
	\maketitle
	\newpage
	\section{Introduction}
	In the standard cosmological model, astrophysical and cosmological observations indicate that the observed gravitational dynamics of the Universe cannot be fully accounted for by visible matter alone \cite{Clifton2012,Nojiri2011,Planck2018,Joyce2015,Koyama2016}. Galaxy rotation curves provide one of the most direct tests of gravity on galactic scales. In Newtonian gravity, the rotational velocity of stars and gas is expected to decrease at large radii once the enclosed baryonic mass becomes approximately constant. Contrary to this expectation, observations show that rotational velocities do not exhibit a Keplerian decline but instead remain nearly constant over extended galactic scales \cite{Rubin1980,Bosma1981,Sofue2001}. Evidence for additional gravitational effects also emerges from galaxy cluster observations \cite{Zwicky1933,Clowe2006,Allen2011}, gravitational-lensing measurements \cite{Bartelmann2001,Hoekstra2008,Massey2010}, and the formation 
	of large-scale cosmic structures \cite{Planck2018,Tegmark2004,Cole2005}. Within the standard cosmological model, these phenomena are commonly interpreted as evidence for an additional non-baryonic dark matter component \cite{Bertone2005,Feng2010}. Despite its phenomenological success, however, dark matter has not yet been directly detected through non-gravitational interactions \cite{Akerib2017,Cui2017,Agnese2018}. This situation has motivated extensive investigations of modified theories of gravity, in which deviations from General Relativity on astrophysical and cosmological scales can reproduce some of the effects commonly attributed to dark matter without introducing additional matter species \cite{Clifton2012,Copeland2006,Tsujikawa2010,FujiiMaeda,Joyce2015,Koyama2016}.
	
	Several theoretical approaches have been proposed as alternatives to the standard cosmological model to explain gravitational phenomena on galactic and cosmological scales. Among them, Modified Newtonian Dynamics (MOND) attributes the observed flattening of galaxy rotation curves to a phenomenological modification of the gravitational acceleration in the low-acceleration regime rather than to particle dark matter \cite{Milgrom1983}. Its relativistic extension, Tensor-Vector-Scalar (TeVeS) gravity, incorporates additional scalar and vector fields to reproduce the MOND phenomenology within a covariant framework \cite{Bekenstein2004}. Another widely studied approach is Scalar-Tensor-Vector Gravity, which extends General Relativity by introducing additional scalar and vector degrees of freedom that modify the effective gravitational interaction \cite{Moffat2006}. A different approach is provided by Hybrid metric-Palatini gravity, which
	combines the metric and Palatini formalisms to provide an alternative
	geometric description of gravity that can account for astrophysical
	observations without invoking a dominant dark matter component
	\cite{Harko2012,Capozziello2015}. Hyperconical Modified Gravity (HMG)
	provides another geometric approach, in which MOND-like galactic dynamics
	are related to the hyperconical cosmological geometry \cite{Monjo2023}.
    An interesting conceptual comparison with the present model arises from the role of background effects in gravitational dynamics. In the present
	framework, the constant background value $K(0)=K(\phi_0,0)$ generates a
	perturbative curvature correction in the weak-field metric, whose relative
	importance increases toward intergalactic scales. The physical mechanisms
	are, however, different. In HMG, the MOND-like galactic behavior originates
	from the hyperconical cosmological geometry, whereas in the present model
	the galactic-scale modifications are governed by the scalar-field dynamics
	and the Vainshtein mechanism. The $K(0)$ term instead provides an additional
	background-curvature contribution that becomes relatively more important
	at larger distances. Conformal gravity represents a further geometric modification of gravity, based on local conformal symmetry, and modifies the gravitational potential in a way that can account for galactic rotation curves without requiring particle dark matter \cite{Mannheim1989,Mannheim2012}. Within the broader class of scalar-tensor theories, Brans-Dicke theory represents the simplest extension of General Relativity through a nonminimally coupled scalar field \cite{BransDicke}, whereas Horndeski theory provides the most general scalar-tensor framework leading to second-order field equations and is characterized by four arbitrary functions of the scalar field and its kinetic term \cite{Horndeski}. Various subclasses of Horndeski theory have subsequently been investigated, particularly cubic Horndeski (Galileon) models, whose nonlinear derivative interactions give rise to the Vainshtein screening mechanism, thereby allowing modifications of gravity on large scales while remaining consistent with local gravitational tests \cite{Nicolis2009,Deffayet2009,Koyama2016}. Alternative scalar-field models have also been proposed in which the scalar field itself constitutes the dark matter component rather than modifying the gravitational interaction \cite{Matos2000,Boehmer2007,Harko2011,Suarez2014,Hui2017,Berezhiani2015}. In addition, viable $f(R)$ gravity models employing screening mechanisms have been widely explored as geometric extensions of General Relativity, allowing deviations from General Relativity on large scales while satisfying local gravitational constraints \cite{HuSawicki2007,Sotiriou2010,DeFelice2010}. Although based on different theoretical principles, these models provide alternative descriptions of gravitational interactions on astrophysical scales and motivate further investigations of weak-field gravitational dynamics. 
	
	In contrast to phenomenological approaches such as MOND, vector-tensor
	theories including STVG, and geometric extensions such as Hybrid
	metric-Palatini gravity, the present framework is based on a covariant
	scalar-tensor theory containing nonlinear scalar derivative interactions.
	Scalar-field modifications of weak-field gravity, Yukawa-type corrections
	associated with massive scalar fields, and screening through nonlinear
	interactions have been extensively studied in scalar-tensor theories.
	Likewise, Horndeski gravity has been investigated in a broad range of
	contexts, including cosmology, gravitational-wave propagation, black-hole
	solutions, Solar System tests, and screening mechanisms
	\cite{Babichev2013,Kobayashi2019,Ezquiaga2017,Copeland2019,HaticeGW,Hatice3}. In particular, previous studies have examined gravitational-wave
	propagation in the presence of a scalar potential or a cosmological
	constant, including the effects associated with the resulting background
	curvature \cite{HaticeGW,HaticeBDGW}. At the same time, gravitational-wave observations have strongly constrained Horndeski theories\cite{Ezquiaga2017,Copeland2019}, motivating the investigation of the astrophysical implications of models that remain compatible with these observations. The present work addresses this question within a gravitational-wave compatible cubic Horndeski model, where nonlinear scalar interactions can affect galactic dynamics through the Vainshtein mechanism while the scalar and background contributions remain relevant to the behavior at larger distances.

	Among scalar-tensor theories, Horndeski theory represents the most general framework leading to second-order equations of motion and is characterized by four arbitrary functions \cite{Horndeski,Deffayet2011,Kobayashi2019}. Owing to its generality, the theory exhibits a rich phenomenology at both cosmological and astrophysical scales. However, large regions of its parameter space are constrained by current observations \cite{Koyama2016,Ezquiaga2017,Baker2017}. In particular, the multimessenger detection of GW170817 established that the propagation speed of gravitational waves is equal to the speed of light to within approximately one part in $10^{15}$ \cite{Abbott2017GW,Abbott2017GRB}. Within the general Horndeski framework, derivative couplings associated with the functions $G_4(X)$ and $G_5(\phi,X)$ can modify the propagation speed of tensor modes and are therefore severely constrained by these observations \cite{Ezquiaga2017,Copeland2019}. Additional constraints from Solar-System experiments and weak-field tests further restrict the phenomenologically viable parameter space \cite{Will2014,Will2, Will3,  Alsing2012,Perivolaropoulos2010,Hatice3}.
	In the present work, we consider the gravitational-wave compatible cubic Horndeski subclass defined by $G_3(X)=\beta X$, $G_{3,\phi}=0$, $G_5=0$, and $G_4=G_4(\phi)$. This subclass satisfies the condition $c_T=1$, ensuring consistency with current gravitational-wave observations, while retaining the simplest nonlinear derivative scalar interaction through the cubic Horndeski term. At the same time, the higher-order derivative couplings associated with $G_{4X}$ and $G_5$ are absent. Unlike the standard covariant cubic Galileon, where the gravitational coupling is typically assumed to be constant, the present framework adopts a scalar-dependent function $G_4(\phi)$. Consequently, this gravitational-wave compatible cubic Horndeski subclass provides a consistent framework for investigating weak-field gravitational phenomena while remaining compatible with current observational constraints.
	On galactic scales, where the gravitational potential satisfies $|\Phi|\ll1$, the weak-field expansion provides an appropriate framework for describing galactic dynamics \cite{Babichev2013,Koyama2016}. Accordingly, the metric perturbations are treated at linear order, while the scalar-field equation is solved by retaining the leading nonlinear derivative interaction \cite{Kimura2012,Brax2012,Andrews2013}. Although the metric and scalar perturbations remain weak, the nonlinear derivative self-interactions of the scalar field can become significant near the transition scale and lead to a scale-dependent suppression of the effective scalar interaction. The resulting framework allows the screened, transition, and asymptotic regimes of the scalar interaction to be analyzed within a consistent weak-field description.
	
	In the present work, we investigate the weak-field solutions of a gravitational-wave compatible subclass of Horndeski gravity containing a dynamical scalar field governed by a nontrivial potential \cite{Hatice2,HaticeGW,HaticeBDGW, Hatice3}. In the weak-field regime, the minimum of the scalar-field potential generates a local de Sitter-like curvature correction. As in scalar-tensor theories with a potential, this contribution behaves as an effective cosmological constant at the linearized level, although this interpretation is valid only within the adopted weak-field expansion, where
	\begin{equation}
		\frac{|K(0)|r^2}{G_4(0)} \ll 1.
	\end{equation}
   The weak-field effects associated with a scalar potential and the resulting background curvature have previously been investigated in scalar-tensor theories, including their implications for Solar System dynamics and gravitational-wave propagation \cite{HaticeGW,HaticeBDGW,Hatice3,Hatice2}. The K(0) contribution is negligible on galactic scales but becomes increasingly important at large distances because of its different radial dependence. Thus, although it does not significantly affect the galactic dynamics considered here, its relative importance increases as the localized attractive contributions decrease with distance. It is therefore retained throughout the analysis to investigate both galactic dynamics and the transition to the intergalactic regime, where it may balance and eventually dominate the attractive gravitational interaction. Consequently, the scalar potential determines the weak-field dynamics through the scalar-field mass and the asymptotic behavior through the effective curvature scale.
	
	In this study, the modifications of weak-field gravitational dynamics induced by nonlinear scalar interactions are analyzed on galactic and intergalactic scales. Particular attention is devoted to the transition region associated with the Vainshtein scale, where the nonlinear scalar contribution becomes comparable to the Newtonian gravitational term and generates the largest departures from the asymptotic behavior. We derive the corresponding rotation curves and investigate the resulting modifications of rotational velocities near the transition scale. We further analyze the radial acceleration relation, the effective-potential structure and stability of circular orbits, and the lensing-dynamical mass difference generated by gravitational slip. On intergalactic scales, the turnaround radius of gravitationally bound systems is derived, and the asymptotic regime is investigated, where the scalar contribution associated with the localized source becomes Yukawa suppressed while the curvature contribution proportional to K(0) governs the large-distance behavior of the gravitational field. A distinctive aspect of the present work is that the galactic and intergalactic regimes are investigated within the same gravitational-wave compatible cubic Horndeski model. This allows the different contributions to the gravitational dynamics to be followed across distance scales. On galactic scales, nonlinear scalar interactions become important around the Vainshtein transition, whereas outside the	Vainshtein region the scalar field approaches the linear regime. At	larger intergalactic distances, the localized scalar contribution becomes Yukawa suppressed and the background-curvature contribution associated
	with $K(0)$ becomes increasingly important.

	This paper is organized as follows. In Sec. II, the gravitational framework is introduced and the weak-field  equations for the gravitational-wave compatible cubic Horndeski model are derived. In Sec. III, the corresponding weak-field solutions are obtained, and their behavior in the screened, transition, and asymptotic regimes is analyzed. In Sec. IV, the galactic implications of the model are investigated, including rotation curves, effective potentials, circular-orbit properties, radial acceleration relations, and the lensing-dynamical mass difference. In Sec. V, the asymptotic intergalactic regime is examined, and the turnaround radius and the large-distance behavior of the gravitational interaction are investigated. Finally, the main results are summarized and their physical implications are discussed in Sec. VI.
	
	\section{Weak Field Equations and Solutions}
	Horndeski theory is the most general four-dimensional scalar-tensor theory yielding second-order field equations for both the metric and the scalar field. The action is given by \cite{Horndeski}
	\begin{equation}
		S= \int d^4x \sqrt{-g}(L_2+L_3+L_4+L_5)+\int dx^4 \sqrt{-g} L_m,
	\end{equation}
	where Horndeski lagrangian densities are,
	\begin{align}
		L_2&= K(\phi,X), \quad L_3= -G_3(\phi, X) \Box\phi, \nonumber \\
		L_4&= G_4(\phi,X) R + G_{4,X} \left[(\Box\phi)^2 - (\nabla_\mu\nabla_\nu\phi)(\nabla^\mu\nabla^\nu\phi)\right], \label{Hornactions} \\
		L_5&= G_5(\phi,X) G_{\mu\nu} \nabla^\mu\nabla^\nu\phi - \frac{1}{6}G_{5,X} \left[(\Box\phi)^3 
		- 3(\Box\phi)(\nabla_\mu\nabla_\nu\phi)(\nabla^\mu\nabla^\nu\phi) 
		+ 2(\nabla^\mu\nabla_\alpha\phi) (\nabla^\alpha\nabla_\beta\phi) (\nabla^\beta\nabla_\mu\phi) \right]. \nonumber
	\end{align}  
	Here, $K$, $G_3$, $G_4$, and $G_5$ are arbitrary functions of the scalar field $\phi$ and its canonical kinetic term $X=-\frac{1}{2}\nabla_\mu\phi\nabla^\mu\phi$, where $\nabla_\mu$ denotes the covariant derivative associated with the metric $g_{\mu\nu}$. The symbol $G_{\mu\nu}$ denotes the Einstein tensor, $g$ is the determinant of the metric tensor, and the Einstein summation convention is assumed throughout this paper.
	
	In this subsection, we derive the weak-field equations for the gravitational-wave compatible cubic Horndeski theory. Following the constraint imposed by the GW170817 event, gravitational waves are required to propagate at the speed of light \(c_T=1\). This condition restricts the Horndeski action to $G_{4X}=0$, $G_5=0$, implying that $G_4$ depends only on the scalar field, $G_4= G_4(\phi)$. While the linearized tensor equations follow the standard weak-field Horndeski formalism, the scalar-field equation is modified by the nonlinear scalar interactions arising from the scalar-dependent coupling $G_4(\phi)$.
	Within this class, the nonlinear scalar self-interactions governed by \(G_3(\phi,X)\) become significant near massive sources, giving rise to the Vainshtein mechanism and suppressing deviations from General Relativity. To derive the weak-field equations describing the gravitational field generated by localized sources, the metric and scalar field are expanded around a Minkowski background as
\begin{align}
	g_{\mu\nu} &= \eta_{\mu\nu} + h_{\mu\nu},\quad  g^{\mu\nu}=\eta^{\mu\nu}-h^{\mu\nu}, \\
\phi(x^\mu) &= \phi_0 + \varphi(x^\mu).
\end{align}
 with $|h_{\mu\nu}|\ll 1$. Here, $\eta_{\mu\nu}$ denotes the Minkowski
 metric, while $\phi_0$ is the constant background value of the scalar
 field. The quantities $h_{\mu\nu}$ and $\varphi(x^\mu)$ are assumed to be
 small perturbations. We next consider the function $K(\phi,X)$ appearing in the action, which contains both kinetic and potential contributions. In the weak-field regime, the kinetic term \(X\) is assumed to remain small, so that higher-order kinetic contributions become subleading. Consequently, \(K(\phi,X)\) may be approximated by \(K(\phi)\). Although the weak-field expansion is performed around a locally Minkowskian background, the constant contribution K(0) is allowed to remain nonvanishing. In the present analysis, this term is not interpreted as defining an exact de Sitter background, but rather as generating perturbative curvature corrections within the weak-field regime. Accordingly, the analysis is restricted to scales satisfying $\frac{|K(0)|r^2}{G_4(0)} \ll 1$, such that the spacetime can still be approximated locally by Minkowski geometry. Within this perturbative treatment, the constant contribution K(0) is retained only through its leading-order background curvature correction proportional to $K(0)r^2$. Mixed terms involving both the background-curvature scale and perturbative fluctuations, such as $K(0)h_{\mu\nu}$ and $K(0)\phi$, are consistently treated as higher-order quantities and are therefore neglected in the linearized approximation. Similar perturbative treatments have been employed in weak-field analyses in the presence of cosmological background contributions \cite{Bernabeu, Bernabeu2011}. In the following, we restrict attention to $K(0)<0$. As shown in our previous analysis of linearized Horndeski theory with a potential, this sign choice corresponds to a de Sitter-like background-curvature contribution, with $K(0)\rightarrow -2\Lambda_{\rm cos} G_4(0)$ in the General Relativity limit. Consequently, the associated large-distance contribution to the gravitational field is repulsive.
	
	The scalar potential \(K(\phi)\) determines the effective scalar mass and therefore the interaction range of the scalar field. Through the associated Compton wavelength \(m_\phi^{-1}\), it controls the transition between the long-range and Yukawa-suppressed regimes. Expanding around the background value, the potential is written as
	\begin{eqnarray}\label{Kexpansion}
		K(\phi)&=&K(0)+K_{,\phi}(0) \varphi+\frac{1}{2}K_{,\phi\phi}(0)\varphi^2+\ldots,\\
		K_{,\phi}(\phi)&=&K_{,\phi}(0)+K_{,\phi\phi}(0)\varphi+\ldots,\nonumber
	\end{eqnarray}
	For notational simplicity, we define $K(\phi,0)\equiv K(\phi)$ and $K(0)\equiv K(\phi_0,0)$. The weak-field equations are then derived under the assumptions
	\begin{equation}
		K(0)\neq 0,
		\qquad
		K_{,\phi}(0)=0.
	\end{equation}
	The condition $K_{,\phi}(0)=0$ implies that the scalar field is expanded around an extremum of the effective potential, ensuring a stable background configuration under small perturbations. The quadratic term $K_{,\phi\phi}(0)$ then determines the effective scalar mass and interaction range, while the constant term K(0) introduces a nonvanishing background contribution that can affect the large-distance weak-field behavior. To simplify the tensor field equations, we introduce the auxiliary field
	\begin{equation}
		h_{\mu\nu}
		=
		\theta_{\mu\nu}
		-\tfrac{1}{2}\eta_{\mu\nu}\theta
		-\eta_{\mu\nu}\sigma \varphi,
	\end{equation}
	where
	\begin{equation}
		\sigma
		=
		\frac{G_{4,\phi}(0)}
		{G_4(0)} .
	\end{equation}
	The parameter $\sigma$ measures the strength of the coupling between the scalar perturbation and the metric perturbations in the linearized field equations. We also impose the Lorentz gauge condition
	\begin{equation}
		\partial^\mu \theta_{\mu\nu}=0 .
	\end{equation}
	Varying the action in Eq.(2) with respect to the metric and	using the weak-field expansions in Eqs.(4) and (6), the tensor field equations reduce, at linear order, to
	\begin{equation}
		\Box \theta_{\mu\nu} = -\frac{16\pi T_{\mu\nu}}{G_4(0)} - \frac{K(0)\eta_{\mu\nu}}{G_4(0)}.
	\end{equation}
	The first term is sourced by the matter energy-momentum tensor, whereas the second arises from the nonvanishing value of the scalar potential at the background scalar field value. Under static spherical symmetry, the scalar perturbation $\varphi(x^\mu)$ depends only on the radial coordinate, so that $\varphi(x^\mu)=\varphi(r)$. Within the weak-field regime, a hybrid approach is adopted in which the tensor equations are treated at the linear level, while the nonlinear derivative interactions of the scalar field are retained. Both the metric and scalar perturbations are assumed to remain in the weak-field regime. However, even when the scalar perturbation itself is small, its derivative self-interactions can become comparable to the linear scalar contribution near massive sources. Under static spherical symmetry, the scalar-field equation contains nonlinear structures of the form
	$$(\Box \varphi)^2-(\nabla_\mu\nabla_\nu\varphi)(\nabla^\mu\nabla^\nu\varphi),$$
	which arise from the $G_3$ function. These terms involve nonlinear combinations of second derivatives while preserving the second-order field equations. As a result, the scalar-field equation cannot in general be linearized, even though the metric and scalar perturbations remain weak. The derivative self-interactions become particularly important near the Vainshtein scale, where their contribution reaches the same order as the linear scalar term and gives rise to the screening mechanism. Retaining these nonlinear terms therefore allows the transition between the screened and asymptotic regimes to be described within the weak-field framework. Under these assumptions, the scalar field equation obtained by varying
	the action in Eq.(2) with respect to $\phi$ reduces to
\begin{equation} 
	\Big[ 
	K_{,X}(0) 
	-2G_{3,\phi}(0) 
	+3\,\frac{G_{4,\phi}^{2}(0)}{G_{4}(0)} 
	\Big]\Box\varphi 
	+K_{,\phi\phi}(0)\varphi 
	+\beta\Big[ 
	(\Box\varphi)^2 
	-(\nabla_{\mu}\nabla_{\nu}\varphi)(\nabla^{\mu}\nabla^{\nu}\varphi) 
	\Big] 
	= 
	2\sigma K(0)-8\pi\sigma \rho .
\end{equation}
	The cubic scalar self-interaction is described by the Horndeski function $G_3(\phi,X)$. Throughout this analysis, it is restricted to the purely kinetic form
	\begin{equation}
		G_3(X)=\beta X,
	\end{equation}
	so that $G_{3,\phi}(0)=0$. The parameter $\beta$ determines the strength of the nonlinear derivative self-interaction of the scalar field. It is convenient to express this coefficient in terms of the nonlinear interaction scale $\Lambda$ through
	\begin{equation}
		\beta=\frac{2}{\Lambda^3},
	\end{equation}
	where $\Lambda$ characterizes the scale at which the nonlinear scalar interaction becomes important.
	The effective kinetic coefficient is defined as
	\begin{equation}
		K_{\rm eff}
		=
		K_{,X}(0)
		+
		\frac{3G_{4,\phi}^2(0)}{G_4(0)} .
	\end{equation}
	The quantity \(K_{\rm eff}\) determines the effective kinetic strength of the scalar degree of freedom in the weak-field regime, while \(K_{,\phi\phi}(0)\) determines the effective scalar mass and therefore the range of the scalar interaction. For convenience, the following definitions are introduced
	\begin{equation}
		m_\phi^2
		\equiv
		-\frac{K_{,\phi\phi}(0)}{K_{\rm eff}},
		\qquad
		\chi
		\equiv
		\frac{\sigma}{K_{\rm eff}},
		\qquad
	\end{equation}
	Here, $m_\phi$ denotes the effective mass of the scalar perturbation and the analysis assumes $m_\phi^2>0$, corresponding to a stable Yukawa-type scalar mode. For $K_{\rm eff}>0$, this requires $K_{,\phi\phi}(0)<0$, thereby avoiding tachyonic instabilities in the weak-field regime. The condition $K_{\rm eff}>0$ is further assumed to avoid ghost-like kinetic instabilities of the scalar degree of freedom. The scalar-field mass also determines the effective interaction range of the scalar force through the associated Compton wavelength \(m_\phi^{-1}\). Consequently, it controls the transition between the long-range regime \((m_\phi r \ll 1)\), where the scalar interaction remains unsuppressed, and the Yukawa regime \((m_\phi r \gg 1)\), where the scalar contribution becomes exponentially suppressed. The dimensionless parameter governing the nonlinear derivative interaction is defined as
	\begin{equation}
		\alpha \equiv \frac{\beta}{K_{\rm eff}}.
	\end{equation}
	Accordingly, the scalar-field equation (12) takes the form
	\begin{equation}
		(\Box - m_\phi^2)\,\varphi
		=
		2\chi\,K(0)
		-
		\alpha\left[
		(\Box\varphi)^2
		-
		(\nabla_\mu\nabla_\nu\varphi)
		(\nabla^\mu\nabla^\nu\varphi)
		\right]-	8\pi\,\chi\,\rho
	\end{equation} 
	Here, $\chi$ is the effective scalar coupling parameter. It determines how efficiently matter sources the scalar field and therefore controls the strength of the scalar response to a given matter distribution. The parameter $\alpha$ sets the strength of the nonlinear derivative self-interaction of the scalar field.
	For convenience, we introduce the effective source term and the nonlinear operator as
	\begin{equation}
		S_{\rm eff}
		=2\chi\,K(0)-8\pi\,\chi\rho,
		\qquad
		\mathcal{I}[\varphi]
		=
		(\Box\varphi)^2
		-
		(\nabla_\mu\nabla_\nu\varphi)
		(\nabla^\mu\nabla^\nu\varphi).
	\end{equation}
	The scalar-field equation (18) then takes the compact form
	\begin{equation}
		(\Box - m_\phi^2)\,\varphi + \alpha\,\mathcal{I}[\varphi]
		=S_{\rm eff}.
	\end{equation}
	This equation represents the weak-field scalar equation of the Horndeski theory, obtained by linearizing the tensor equations around the Minkowski background while retaining the leading nonlinear derivative interaction in the scalar field. The operator $(\Box - m_\phi^2)$ describes the propagation of a massive scalar degree of freedom,  whereas the source term $8\pi\chi\rho$ determines its coupling to matter. The additional contribution proportional to $K(0)$ arises from the non-vanishing background value of the function $K$ and acts as an effective vacuum source. The nonlinear term $\alpha\,\mathcal{I}[\phi]$ corresponds to a Galileon-type self-interaction responsible for the Vainshtein screening mechanism. Depending on the physical regime, different terms dominate the scalar dynamics. Near massive sources, the nonlinear derivative interactions can become comparable to or larger than the linear scalar contribution, leading to the suppression of scalar-mediated effects. At sufficiently large distances, the nonlinear derivative interactions become subdominant and the scalar field approaches the asymptotic solution governed by the linearized dynamics.
	
   For the galactic applications considered below, the matter source is modeled
   as an extended, nonrelativistic pressureless distribution with energy-momentum
   tensor 
    \begin{equation}
    	T_{\mu\nu}=\rho(r)u_\mu u_\nu ,
    \end{equation}
    which, in the static weak-field limit, gives
    $T_{00}\simeq\rho(r)$, while $T_{0i}=0$ and $T_{ij}\simeq0$. The corresponding
    enclosed mass is defined by
    \begin{equation}
    	M(r)=4\pi\int_0^r \rho(r')\,r'^2\,dr' .
    \end{equation}
    Within the weak-field approximation considered here, we further assume that the enclosed mass varies slowly,
    \begin{equation}
    	M'(r)\ll\frac{M(r)}{r}.
    \end{equation}
    Under this approximation, $M(r)$ can be treated locally as approximately
    constant at leading order. The tensor equation in Eq.(11) can then be solved
    under the Lorenz gauge condition in Eq.(10) in terms of the enclosed mass
    $M(r)$. 
    \begin{align}
    	\theta_{00} &\simeq
    	\frac{4M(r)}{G_4(0)r}
    	+\frac{K(0)}{6G_4(0)}r^2,
    	&
    	\theta_{xx} &=
    	-\frac{K(0)}{4G_4(0)}(y^2+z^2),
    	\\
    	\theta_{yy} &=
    	-\frac{K(0)}{4G_4(0)}(x^2+z^2),
    	&
    	\theta_{zz} &=
    	-\frac{K(0)}{4G_4(0)}(x^2+y^2).
    \end{align}
    These tensor solutions determine the metric perturbations together with the
    scalar-field solution through the auxiliary-field relation in Eq.(8).
    The corresponding scalar-field solutions are derived in the following
    subsection.
	\section{Scalar field solutions in a static, spherically symmetric background}	
	The weak-field equations derived in the previous subsection are now solved. Whereas the tensor solutions coincide with those obtained in the standard linearized Horndeski theory, the scalar-field solution is modified by the scalar-dependent coupling $G_4(\phi)$ and the nonlinear cubic interaction. This solution forms the basis for the subsequent weak-field analyses presented in this work. To obtain the corresponding weak-field solution, we consider a static, spherically symmetric galactic background. The analysis is performed in the weak-field limit, with the matter source described by an extended mass distribution appropriate for galactic systems. Static spherically symmetric solutions in the linear regime, including
	the effects of a cosmological background on local gravitational dynamics
	\cite{Bernabeu, Bernabeu2011}, as well as screened solutions and the transition between
	screened and unscreened regimes in Galileon and Horndeski theories
	\cite{Nicolis2009,Kase2013,Babichev2010,Vainshtein1972,Kimura2012,	Kobayashi2015,Babichev2013,Andrews2013}, have been extensively investigated. While these studies primarily focused on compact sources and scalar-field screening effects, the present analysis focuses on the weak-field scalar dynamics relevant to galactic systems and their connection with the outer intergalactic regime. Under these assumptions, the scalar field depends only on the radial coordinate, with the matter source entering through the enclosed mass \(M(r)\). The presence of nonlinear derivative interactions, together with the scalar potential K($\phi$), makes the scalar dynamics intrinsically nonlinear. As a consequence, the Vainshtein mechanism suppresses the fifth force in dense regions, while the scalar interaction becomes effectively unscreened at large distances. To describe the galactic background, a static and spherically symmetric spacetime is considered, described by the Schwarzschild-like line element
	\begin{equation}
		ds^2 = -A(r)dt^2 + B(r)dr^2 + r^2 d\Omega^2 ,
	\end{equation}
	where the metric functions depend only on the radial coordinate. In the weak-field regime, the metric functions are expanded as
	\begin{equation}
		A(r)=1+2\Phi(r), 
		\qquad 
		B(r)=1-2\Psi(r),
	\end{equation}
	with $|\Phi| \ll 1$ and $|\Psi| \ll 1$.
For the spherically symmetric configuration considered here, the matter source enters the scalar-field solution through the enclosed mass M(r) defined in Eq.(22). Under the assumption of spherical symmetry, the scalar perturbation depends  only on the radial coordinate, $\varphi=\varphi(r)$, and the scalar-field  equation given in Eq.(20) then takes the form
  \begin{equation}
	K_{\rm eff}\left(\varphi''+\frac{2}{r}\varphi'\right)
	+K_{,\phi\phi}(0)\varphi
	+\beta\left(\frac{4}{r}\varphi'\varphi''
	+\frac{2}{r^2}(\varphi')^2\right)
	=
	2\sigma K(0)-8\pi\sigma \rho .
  \end{equation}
This equation can equivalently be written in the total-derivative form 
	\begin{equation}
		\frac{1}{r^2}\frac{d}{dr}
		\left[
		K_{\rm eff}r^2\varphi'
		+2\beta r(\varphi')^2
		\right]
		+K_{,\phi\phi}(0)\varphi
		=
		2\sigma K(0)-8\pi\sigma \rho
	\end{equation}
	The scalar field Eq.(29) does not yield a single analytic solution valid across all radial scales due to the combined effects of nonlinear derivative interactions and the scalar-field mass. Similar regime-based analyses have been widely employed in studies of Vainshtein screening and Galileon/Horndeski theories \cite{Nicolis2009,Babichev2013,Koyama2016, DeFelice2010}. As a result, different radial regimes arise in a static, spherically symmetric background. Around the Vainshtein radius ($r \sim r_V$), the system enters a transition regime in which nonlinear screening effects become less efficient, and the scalar field evolves from the screened to the unscreened regime. 
	
	At intermediate scales ($r_V \ll r \ll m_\phi^{-1}$), the nonlinear derivative interaction becomes negligible, while the scalar mass term remains subdominant. The scalar-field dynamics is therefore approximately governed by the linearized field equation. In this regime, the scalar force can contribute significantly to galactic dynamics. At sufficiently large radii, ($r \gg m_{\phi}^{-1}$), the scalar potential induces an effective mass and leads to an exponential Yukawa suppression of the field. These regimes must therefore be analyzed separately to understand the galactic-scale behavior of the scalar field.
	We begin with the screened regime, defined by $r \ll r_V$, where the nonlinear derivative term satisfies
	$2\beta r(\varphi')^2 \gg K_{\rm eff}r^2\varphi'$. Retaining the contribution from the scalar potential, the scalar field equation in Eq.(29) reduces to
	\begin{equation}
		\frac{1}{r^2}\frac{d}{dr}
		\left[
		2\beta r(\varphi')^2
		\right]
		+
		K_{,\phi\phi}(0)\varphi
		\simeq
		2\sigma K(0)-8\pi\sigma \rho .
	\end{equation}
	Multiplying the equation by $r^2$ integrating from the origin to r, and expressing the matter source in terms of the enclosed mass function M(r), we obtain
	\begin{equation}
		2\beta r(\varphi')^2
		+
		K_{,\phi\phi}(0)
		\int_0^r \varphi(\tilde r)\tilde r^2 d\tilde r
		\simeq
		\frac{2\sigma K(0)}{3}r^3
		-2\sigma M(r) .
	\end{equation}
	The scalar field derivative can be formally written as
	\begin{equation}
		\varphi'(r)
		\simeq
		\pm
		\left\{
		\frac{1}{2\beta r}
		\left[
		\frac{2\sigma K(0)}{3}r^3
		-2\sigma M(r)
		-
		K_{,\phi\phi}(0)
		\int_0^r \varphi(\tilde r)\tilde r^2 d\tilde r
		\right]
		\right\}^{1/2}.
	\end{equation}
	At this stage, both signs correspond to the two formal branches of the scalar-field solution. In the following, we restrict our attention to the branch corresponding to $\sigma<0$, which leads to an attractive scalar interaction and is used consistently throughout the analyses of the dynamical mass and the circular velocity. The sign of $\varphi'(r)$ is then fixed by requiring continuity with the linear solution.
	
	In this regime, the nonlinear derivative interaction dominates over the linear kinetic term. For $r \ll m_\phi^{-1}$, the scalar mass term proportional to $K_{,\phi\phi}(0)$ is negligible, and the corresponding scalar force is Vainshtein suppressed. Keeping the background contribution for the moment, the scalar field equation in Eq.(31) reduces to
	\begin{equation}
		2\beta r(\varphi')^2
		\simeq
		\frac{2\sigma K(0)}{3}r^3
		-2\sigma M(r)
	\end{equation}
	which yields
	\begin{equation}
		\varphi'_{\rm scr}(r)
		\simeq
		\left[
		\frac{\sigma}{\beta r}
		\left(
		\frac{K(0)}{3}r^3
		-
		M(r)
		\right)
		\right]^{1/2}.
	\end{equation}
	On galactic scales, the background contribution from the scalar potential is typically negligible compared to the local matter source, since $|K(0)|r^3 \ll M(r)$. The screened field equation in Eq.(33) therefore further reduces to
	\begin{equation}
		2\beta r(\varphi')^2
		\simeq
		-2\sigma M(r).
	\end{equation}
	The scalar-field gradient in Eq.(34) then becomes
	\begin{equation}
		\varphi'_{\rm scr}(r)
		\simeq
		\left[
		-\frac{\sigma M(r)}{\beta r}
		\right]^{1/2}.
	\end{equation}
	For $\beta>0$, the reality of the scalar-field solution requires the argument of the square root to be positive. Since $M(r)>0$ and $r>0$, this implies $\sigma<0$. 
	Thus, the scalar dynamics are controlled by the nonlinear derivative interaction responsible for the screening mechanism. This behavior is characteristic of Vainshtein-screened solutions in Galileon-type theories~\cite{Nicolis2009, Deffayet2011, Kimura2012}. In the linear regime, where the nonlinear derivative interaction is negligible, the scalar field equation in Eq.(29) gives
	\begin{equation}
		\varphi'_{\rm lin}(r)
		\simeq
		-\frac{2\sigma M(r)}{K_{\rm eff} r^2}.
	\end{equation}
	The Vainshtein radius is defined as the scale at which the magnitudes of the linear and nonlinear derivative contributions become comparable,
	\begin{equation}
		\left|2\beta r(\varphi')^2\right|
		\sim
		\left|K_{\rm eff}r^2\varphi'\right|.
	\end{equation}
	Using the linear solution at the transition scale together with \(\beta=2/\Lambda^3\), the Vainshtein scale is determined by the magnitude of the scalar coupling,
	\begin{equation}
		r_V^3(r)
		\simeq
		\frac{8|\sigma|M(r)}
		{\Lambda^3K_{\rm eff}^2}.
	\end{equation}
	The absolute value appears because the Vainshtein radius is defined by equating the magnitudes of the linear and nonlinear derivative contributions and therefore represents a positive characteristic length scale. The sign of $\sigma$ affects the sign of the scalar contribution to the gravitational dynamics and will be discussed below in the analysis of the radial acceleration and circular velocity.
	Since the scalar force is determined by the scalar-field gradient, the suppression of \(\varphi'(r)\) directly suppresses the fifth force. In the screened regime, the Vainshtein mechanism ensures that
	\begin{equation}
		|F_\phi(r)| \ll |F_N(r)|.
	\end{equation}
	Compared with the linear scalar-force solution, the screened scalar force scales as
	\begin{equation}
		\frac{F_\phi^{\rm scr}}{F_\phi^{\rm lin}}
		\sim
		\left(\frac{r}{r_V}\right)^{3/2},
	\end{equation}
	which shows that the scalar contribution becomes strongly suppressed for $r \ll r_V$. The screened solution obtained in the inner region is valid only when the nonlinear derivative interaction dominates over the linear kinetic term. As the radial distance increases, the nonlinear contribution weakens, leading the scalar field to transition toward the intermediate regime. In the transition region, both the linear and nonlinear derivative terms must be retained. For $r \ll m_\phi^{-1}$, the scalar mass term remains negligible, while on galactic scales the background contribution proportional to K(0) is subdominant compared with the local matter source. After one radial integration of Eq.(29), the scalar-field equation reduces to
	\begin{equation}
		K_{\rm eff} r^2 \varphi'
		+
		2\beta r(\varphi')^2
		\simeq
		-2\sigma M(r).
	\end{equation}
	Since $\sigma<0$, the source term can equivalently be written as
	$-2\sigma M(r)=2|\sigma|M(r)$. Substituting $\beta=2/\Lambda^3$, the above equation becomes a quadratic equation for $\varphi'$,
	\begin{equation}
		\frac{4}{\Lambda^3}r(\varphi')^2
		+
		K_{\rm eff}r^2\varphi'
		-
		2|\sigma|M(r)
		=0.
	\end{equation}
	Solving this equation for $\varphi'$ gives
	\begin{equation}
		\varphi'(r)
		=
		\frac{K_{\rm eff}\Lambda^3 r}{8}
		\left[
		-1
		\pm
		\sqrt{
			1+
			\frac{32|\sigma|M(r)}
			{\Lambda^3K_{\rm eff}^2r^3}
		}
		\right].
	\end{equation}
	We select the positive branch, which continuously approaches the linear weak-field solution as the nonlinear contribution becomes negligible. Using the Vainshtein radius defined in Eq.(39), the term under the square root can be expressed as
	\begin{equation}
		\frac{32|\sigma|M(r)}
		{\Lambda^3K_{\rm eff}^2r^3}
		=
		4\left(\frac{r_V(r)}{r}\right)^3.
	\end{equation}
	Therefore, the scalar-field gradient in the transition regime can be written as
	\begin{equation}
		\varphi'_{tr}(r)
		=
		\frac{K_{\rm eff}\Lambda^3 r}{8}
		\left(
		-1+
		\sqrt{
			1+4\left(\frac{r_V(r)}{r}\right)^3
		}
		\right).
	\end{equation}
The transition regime, corresponding to distances near the Vainshtein scale, $r \sim r_V$, is particularly relevant because the scalar acceleration reaches the same order of magnitude as the Newtonian acceleration. Consequently, the scalar field contributes to the total acceleration at leading order and cannot be neglected in the description of test-particle motion. The transition solution derived above therefore provides the natural framework for investigating the dynamical effects of the scalar field. Since the Vainshtein radius is determined by the enclosed mass,
	$r_V(r)\propto M^{1/3}(r)$, the location of the transition region
	depends on the matter distribution. In the following transition-region analysis, we focus on the regime where the enclosed mass varies slowly with radius,
	\begin{equation}
		M'(r)\ll \frac{M(r)}{r},
	\end{equation}
	so that the associated Vainshtein scale changes only weakly across the transition region.  This permits a local analysis over a limited radial interval in which $r_V(r)$ may be treated as approximately constant. The solution exhibits the same qualitative screened-to-unscreened transition found in Vainshtein-screened solutions of Galileon and Horndeski theories \cite{Babichev2010,Kimura2012,Kase2013}. Here, the branch is chosen such that the scalar contribution gives an attractive correction to the gravitational force. The solution smoothly connects the screened and linear regimes. In the limit $r \gg r_V$, the nonlinear derivative interaction becomes negligible and one recovers the linear-regime behavior
	$\varphi'(r)\propto \frac{M(r)}{r^2}$.
	In the opposite limit, $r \ll r_V$, the nonlinear derivative interaction dominates and the scalar field behaves as $
	\varphi'(r)\propto \left(\frac{M(r)}{r}\right)^{1/2}$. The transition-region solution therefore connects the screened and linear asymptotic regimes in an approximate manner.
	
	In the intermediate regime, $r_V \ll r \ll m_\phi^{-1}$, the nonlinear derivative interaction becomes negligible, while the scalar mass term can still be ignored since $m_\phi r \ll 1$. The scalar-field equation therefore reduces approximately to a linear Poisson-type equation. Under these conditions, integrating the scalar-field equation in Eq.(29) gives
	\begin{equation}
		\varphi'_{\rm int}(r)
		\simeq
		-\frac{2\sigma M(r)}{K_{\rm eff}r^2}
		+
		\frac{2\sigma K(0)}{3K_{\rm eff}}\,r .
	\end{equation}
	Outside the matter distribution, where $M(r)\simeq M$, a further integration yields
	\begin{equation}
		\varphi_{\rm int}(r)
		\simeq
		+
		\frac{2\sigma M}{K_{\rm eff}r}
		+
		\frac{\sigma K(0)}{3K_{\rm eff}}r^2 .
	\end{equation}
	The first term corresponds to the scalar field generated by the galactic matter distribution, whereas the second term originates from the background contribution associated with K(0). Since $r \ll m_\phi^{-1}$, the scalar mass does not yet modify the radial profile significantly, and the matter-sourced component retains the characteristic long-range behavior of a massless scalar field.
	
	In the outer regime, $r \gg m_\phi^{-1}$, the nonlinear derivative self-interaction terms proportional to $\beta$ are subdominant, and the scalar-field equation reduces to a linear massive equation. In this region, the scalar mass term can no longer be neglected. Using the definition of the effective scalar mass given in Eq.(16), the scalar-field equation in Eq.(29) becomes
	\begin{equation}
		K_{\rm eff}
		\left(
		\varphi''+\frac{2}{r}\varphi'
		-
		m_\phi^2\varphi
		\right)
		=
		2\sigma K(0).
	\end{equation}
	The corresponding outer-region solution is
	\begin{equation}
		\varphi_{\rm out}(r)
		=	\frac{2\sigma M}{K_{\rm eff}r}
		e^{-m_\phi r}+
		\frac{2\sigma K(0)}{K_{,\phi\phi}(0)}.
	\end{equation}
	The solution consists of an asymptotic background value together with a Yukawa-suppressed contribution sourced by the matter distribution. This behavior indicates that the local scalar interaction becomes effectively short-ranged and is exponentially suppressed beyond the Compton wavelength $m_\phi^{-1}$. Consequently, the scalar field generated by the localized matter distribution rapidly decays at sufficiently large distances. The asymptotic solution is therefore dominated by the constant contribution associated with $K(0)$, which may influence the effective background geometry and hence the gravitational dynamics on intergalactic scales. 
	
	Consequently, the validity of the different scalar-field regimes is determined by the hierarchy of scales associated with the Vainshtein radius $r_V$ and the scalar Compton wavelength $m_\phi^{-1}$. The screened regime applies for $r \ll r_V$, where nonlinear derivative interactions dominate and suppress the scalar force. The intermediate regime corresponds to $r_V \ll r \ll m_\phi^{-1}$, where the nonlinear derivative interactions become negligible while the Yukawa suppression remains ineffective. Finally, the outer regime is reached for $r \gg m_\phi^{-1}$, where the scalar interaction becomes Yukawa suppressed.

	\section{Galaxy dynamics}
	In this section, we apply the weak-field solutions derived in the previous section to the study of galactic dynamics within the gravitational-wave compatible cubic Horndeski theory. The following analysis derives the circular velocity, the corresponding radial acceleration relation, the effective potential governing orbital motion, and the lensing-dynamical mass difference, thereby describing how the scalar interaction modifies the dynamics of galaxies. To this end, we analyze the motion of timelike test particles in a static, spherically symmetric spacetime. We work in the Schwarzschild-like coordinates introduced in Eq.(26) and restrict the motion to the equatorial plane by setting $\theta=\pi/2$. The Lagrangian describing the motion of test particles and photons is then given by
	\begin{equation}
		2\mathcal{L}
		=
		-A(r)\dot{t}^{\,2}
		+
		B(r)\dot{r}^{\,2}
		+
		r^2\dot{\psi}^{\,2},
	\end{equation}
	The dot denotes differentiation with respect to proper time for timelike trajectories and an affine parameter for null trajectories. Owing to the time-translational and rotational symmetries of the spacetime, the specific energy E and angular momentum L are conserved. Using these conserved quantities, the radial equation of motion can be written as
	\begin{equation}\label{radialeom}
		\dot{r}^{2}
		=
		\frac{1}{B}
		\left(
		\varepsilon
		+
		\frac{E^{2}}{A}
		-
		\frac{L^{2}}{r^{2}}
		\right),
	\end{equation}
	Here $E$ and $L$ represent the conserved specific energy and angular momentum, respectively. Using these relations together with the metric condition leads to the radial equation	with $\varepsilon=-1$ for timelike particles and $\varepsilon=0$ for photons. Dividing the radial equation in Eq.(53) by $\dot{\psi}^{2}$ gives the corresponding orbital equation.
	\begin{equation}
		\left(\frac{dr}{d\psi}\right)^2
		=
		\frac{\dot{r}^{\,2}}{\dot{\psi}^{\,2}}
		=
		r^4\left(
		\frac{\varepsilon}{L^2 B(r)}
		+
		\frac{E^2}{L^2 A(r)B(r)}
		-
		\frac{1}{B(r)r^2}
		\right).
	\end{equation}
	The orbital equations obtained above provide the basis for analyzing galaxy rotation curves, radial acceleration relations, and the distinction between dynamical and lensing masses.
	\subsection{Galaxy rotation curves}
In this subsection, we investigate the circular motion of timelike test particles in the weak-field regime and derive the corresponding circular velocity. The aim is to derive analytical predictions for the scalar contribution across the
Vainshtein transition, rather than to fit individual galactic rotation curves.

To construct the weak-field metric used in the following analysis, the tensor solutions in Eqs.(24) and (25) and the scalar solution in Eq.(46) are combined through the auxiliary-field relation in Eq.(8). The resulting metric perturbations are then brought into Schwarzschild-like form through a sequence of coordinate transformations. These transformations also introduce the $r\varphi'(r)$ contribution in the radial metric component. A detailed derivation of the coordinate transformations leading to this form was presented in our previous works \cite{Hatice3,HaticeGW}. The resulting Schwarzschild-like line element is
	\begin{equation}
		ds^2 =
		- \left( 1 - \frac{2M(r)}{G_4(0)\,r} + \frac{K(0)}{6G_4(0)}\,r^2 - \sigma \varphi(r) \right) dt^2
		+ \left( 1 + \frac{2M(r)}{G_4(0)\,r} - \frac{K(0)}{6G_4(0)}\,r^2 + \sigma r \varphi'(r) \right) dr^2
		+ r^2 d\Omega^2 \, .
	\end{equation}
	Here, M(r) denotes the enclosed mass, while the last terms in the metric components represent the scalar-field contribution. The term proportional to K(0) corresponds to an additional contribution to the gravitational potential. In the transition regime $(r \ll m_\phi^{-1}$), the scalar-field gradient is described by Eq.(39). Using this result, we now derive the effective gravitational potential and the corresponding rotation curves.
	In the weak-field regime, the metric function $g_{tt}$ can be written in terms of the effective gravitational potential as
	\begin{equation}
		g_{tt}=-(1+2\Phi_{\rm eff}).
	\end{equation}
	For the metric under consideration, this relation yields
	\begin{equation}
		\Phi_{\rm eff}
		=
		-\frac{M(r)}{G_4(0)\,r}
		+
		\frac{K(0)}{12G_4(0)}\,r^2
		-
		\frac{\sigma}{2}\varphi(r).
	\end{equation}
	The tangential velocity of a test particle moving on a circular orbit in the weak-field, static, and spherically symmetric spacetime is given by
\begin{equation}
	v_c^2(r)=\frac{r A'(r)}{2A(r)}.
\end{equation}
	To determine how the scalar interaction modifies the circular-velocity profile, we employ the full transition-region solution for the scalar-field gradient rather than its screened or linear asymptotic approximations. The asymptotic solutions describe only the limiting regimes, whereas the transition solution provides a continuous description of the nonlinear scalar dynamics across the Vainshtein transition. The weak-field solutions and the corresponding transition solution employed in this section follow from the analysis presented in Sec. III. We apply these solutions to calculate the galactic rotation curves and examine the scalar contribution to the circular velocity in the vicinity of the Vainshtein transition. Since our primary interest is the behavior of the circular velocity near the transition scale, we assume that the enclosed mass varies sufficiently slowly over the corresponding radial interval, $M'(r)\ll \frac{M(r)}{r}$.
	This approximation is appropriate when the characteristic scale over which the mass profile changes is much larger than the width of the transition region. Consequently, the contribution proportional to $M'(r)$ in the radial derivative of the effective potential can be neglected. Starting from the weak-field expression for the tangential velocity in Eq.(55) and substituting the transition-region solution for $\varphi'(r)$ given in Eq.(46), we obtain
	\begin{equation}
		v_c^2(r)
		\simeq
		\frac{M}{G_4(0)\,r}
		+
		\frac{K(0)}{6G_4(0)}\,r^2
		-
		\frac{\sigma K_{\rm eff}\Lambda^3 r^2}{16}
		\left[
		-1+
		\sqrt{
			1+4\left(\frac{r_V}{r}\right)^3
		}
		\right] .
	\end{equation}
	The Newtonian contribution reproduces the familiar Keplerian behavior of the circular velocity. On galactic scales, the background-curvature term proportional to K(0) is expected to remain negligible for values of K(0) consistent with the observed cosmological constant. The dominant departure from the Newtonian prediction therefore originates from the nonlinear scalar interaction. Through the scalar-field gradient  $\varphi'(r)$, this contribution is governed by the ratio $r_V/r$, which characterizes the transition between the screened and linear scalar-field regimes.  For the attractive branch ($\sigma<0$), the scalar interaction provides an additional attractive contribution to the gravitational field, leading to a larger circular velocity than in the purely Newtonian case. 
	
	Galaxy rotation curves have also been investigated within covariant cubic Galileon, Yukawa-type, and beyond-Horndeski theories. Covariant cubic Galileon models have been shown to reproduce the observed flattening of galaxy rotation curves through a Vainshtein-screened scalar interaction, while Yukawa-type gravity predicts an exponential correction characterized by a finite interaction length. In beyond-Horndeski theories, the circular velocity receives additional contributions associated with the partial breaking of the Vainshtein mechanism inside extended sources. By contrast, the present analysis is restricted to the transition regime of the gravitational-wave compatible cubic Horndeski theory. In this regime, the scalar contribution enhances the circular velocity relative to the Newtonian prediction and is governed by the nonlinear Vainshtein transition through the ratio $r/r_V$. Therefore, the present result describes the enhancement of the circular velocity in the transition regime rather than the asymptotically flat behavior of galaxy rotation curves. In the limit $\beta\rightarrow0$, the cubic interaction vanishes and the	Vainshtein mechanism is no longer present. Since $\beta=2/\Lambda^3$, this limit gives $r_V\rightarrow0$, and the scalar contribution recovers the $M(r)/r$ dependence of the linearized Horndeski weak-field result \cite{Secuk2025}. Thus, in the absence of the cubic interaction, the result reduces to the corresponding linearized Horndeski behavior.
		
	To gain further insight into the radial dependence of the scalar contribution, we examine its behavior in the screened and linear asymptotic regimes. In the screened regime (\(r \ll r_V\)), the scalar contribution to the circular velocity behaves as $v_c^2(r)\propto r^{1/2}$.
	In the linear regime (\(r_V\ll r\ll m_\phi^{-1}\)), it follows the same radial scaling as the Newtonian term, giving $	v_c^2(r)\propto r^{-1}$. Therefore, neither asymptotic regime by itself produces an exactly flat rotation curve. Consequently, any nontrivial modification of the rotational dynamics is expected to occur in the transition region connecting these two limiting regimes. To investigate this regime more closely, we consider the full transition-region expression for the scalar contribution to the circular velocity. The transition scale associated with the transition can be identified from the stationary-point condition
	\begin{equation}
		\frac{d v_c^2}{dr}\simeq 0.
	\end{equation}
	Using the full transition-region expression, this condition yields
	\begin{equation}
		r_* = 2^{-1/3}r_V
		\simeq
		0.8r_V.
	\end{equation}
	The existence of the stationary point shows that the scalar contribution possesses a characteristic radial scale at which its magnitude is maximized. The corresponding radius is determined entirely by the Vainshtein scale through $r_* = 2^{-1/3}r_V$. This implies that the influence of the scalar interaction is localized around the Vainshtein transition, where its impact on the circular velocity is expected to be greatest. Away from this region, the scalar contribution gradually approaches the screened and linear asymptotic behaviors discussed above. The effects of Vainshtein screening on galactic rotation curves have previously been investigated in cubic Galileon and related scalar-tensor theories \cite{KoyamaSakstein2015, Sakstein2015,Rinaldi2017}. These studies primarily examined the phenomenological impact of the scalar interaction on galactic rotation curves. In contrast, the present analysis analytically identifies the transition radius at which the scalar contribution reaches its maximum. 
	
\begin{figure}[H]
		\centering
		\includegraphics[
		draft=false,
		width=7.5cm
		]{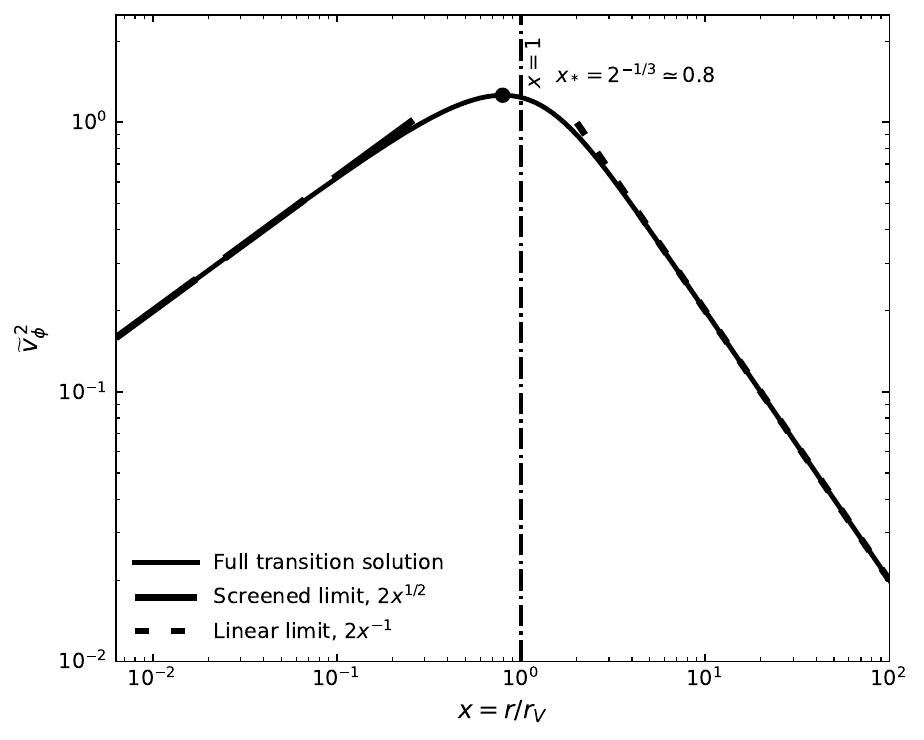}
		\caption{
			Dimensionless scalar contribution to the squared circular velocity,
			$\widetilde{v}_{c}^{\,2}$, as a function of the normalized radius
			$x=r/r_V$. The solid curve denotes the transition solution, while the
			dashed and dotted curves show the screened and linear asymptotic
			limits, respectively. The vertical dash-dotted line marks the
			Vainshtein scale $x=1$, and the filled circle indicates
			$x_*=2^{-1/3}\simeq0.8$, where the scalar contribution reaches its
			maximum.
		}
		\label{fig:vainshtein-transition-scalar-profile}
	\end{figure}

	To illustrate the radial behavior of the scalar contribution across the
	Vainshtein transition, we plot the transition solution together with its
	screened and linear asymptotic limits. For this purpose, we introduce the
	normalized radius $x=\frac{r}{r_V}$, and define the dimensionless scalar contribution by factoring out the model-dependent prefactor, $\widetilde{v}_{c}^{\,2}\equiv-\frac{16\,v_{\phi}^{2}}{\sigma K_{\rm eff}\Lambda^{3}r_V^{2}}$. The resulting dimensionless profile is shown in
	Fig.\ref{fig:vainshtein-transition-scalar-profile}, together with the
	corresponding screened and linear asymptotic limits.
	As shown in Fig.\ref{fig:vainshtein-transition-scalar-profile}, the
	transition solution approaches the screened and linear asymptotic limits
	in the corresponding regimes and exhibits a maximum at
	$x_*=2^{-1/3}\simeq0.8$. This behavior confirms that the scalar
	interaction possesses a characteristic radial scale, determined by the
	Vainshtein radius, at which its contribution to the circular velocity is
	greatest.  However, the scalar contribution does not approach a constant
	value over an extended radial range, and therefore does not by itself
	indicate an extended flat rotation curve.
	
	To investigate the effect of the scalar interaction on the circular velocity over galactic scales, we consider an extended spherical mass distribution described by a Hernquist profile. The Hernquist profile provides a simple
	analytic model of a spherical galaxy in which the enclosed mass approaches
	a finite total mass at large radii. This allows the modification of the
	rotation curve to be followed across the Vainshtein transition, where the
	scalar field changes from the nonlinear screened regime to the outer
	linear regime.
	\begin{equation}
		M(r)=M_{\rm gal}\frac{r^2}{(r+a)^2},
	\end{equation}
	where $M_{\rm gal}$ is the total galactic mass and $a$ is the Hernquist
	scale radius, which sets the characteristic radial scale of the mass
	distribution. Since $M(r)\rightarrow M_{\rm gal}$ at large radii, the
	corresponding radius-dependent Vainshtein scale can be written as
	\begin{equation}
		r_V^3(r)
		=
		r_{V,\infty}^3\frac{r^2}{(r+a)^2},
		\qquad
		r_{V,\infty}^3
		=
		\frac{8|\sigma|M_{\rm gal}}
		{\Lambda^3K_{\rm eff}^2}.
	\end{equation}
	Here, $r_{V,\infty}$ denotes the limiting value of the radius-dependent
	Vainshtein scale $r_V(r)$ as the enclosed mass approaches the total
	galactic mass at large radii. For convenience, we introduce the dimensionless quantities
	\begin{equation}
		x=\frac{r}{a},
		\qquad
		\xi=\frac{r_{V,\infty}}{a},
		\qquad
		v_0^2=\frac{G_NM_{\rm gal}}{a},
	\end{equation}
	where $x$ is the radius normalized by the Hernquist scale radius,
	$\xi$ measures the asymptotic Vainshtein scale relative to the galactic
	scale radius, and $v_0$ defines the characteristic squared velocity of
	the Hernquist system. We further define the dimensionless scalar amplitude
	asand the dimensionless scalar amplitude as $\epsilon_{\phi}\equiv-\frac{\sigma G_4(0)K_{\rm eff}\Lambda^3a^3}{16M_{\rm gal}}$, which parametrizes the strength of the scalar contribution to the
	dimensionless rotation curve. In terms of these quantities, the circular
	velocity becomes
	\begin{equation}
		\frac{v_c^2}{v_0^2}
		=
		\frac{x}{(1+x)^2}
		+
		\epsilon_{\phi} x^2
		\left[
		-1+
		\sqrt{
			1+\frac{4\xi^3}{x(1+x)^2}
		}
		\right].
	\end{equation}
	We consider the branch $\epsilon_{\phi}>0$, for which the scalar contribution
	enhances the circular velocity. For the numerical illustration, it is useful to characterize the relative strength of the scalar contribution, $v_\phi^2$, with respect to the	Newtonian contribution, $v_N^2$. We therefore define
	\begin{equation}
		\delta_{\rm out}
		\equiv
		\lim_{r\rightarrow\infty}
		\frac{v_\phi^2}{v_N^2}
		=
		2\epsilon_{\phi}\xi^3.
	\end{equation}
    The transition radius $r_{\rm tr}$ marks the radial scale at which the local Vainshtein radius becomes equal to the physical radius. It is therefore defined by the condition $r_V(r_{\rm tr})=r_{\rm tr}$, which gives
	\begin{equation}
		\xi^3=x_{\rm tr}(1+x_{\rm tr})^2,
		\qquad
		x_{\rm tr}=\frac{r_{\rm tr}}{a}.
	\end{equation}
	For the illustrative curves below, we take $M_{\rm gal}=10^{11}M_\odot$, $a=3\,{\rm kpc}$, and $r_{\rm tr}=20\,{\rm kpc}$, and consider
	$\delta_{\rm out}=0.1$, $0.3$, and $0.5$. These representative values
	illustrate how the scalar interaction progressively modifies the
	Newtonian Hernquist decline across the transition region.
	\begin{figure}[H]
		\centering
		\includegraphics[
		draft=false,
		width=7.5cm
		]{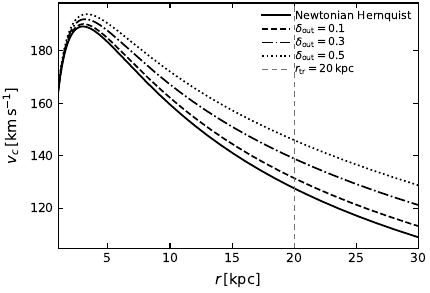}
		
		\caption{
			Circular velocity profiles for a Hernquist mass distribution with
			$M_{\rm gal}=10^{11}M_{\odot}$ and $a=3\,{\rm kpc}$.
			The solid black curve shows the Newtonian Hernquist result, while
			the remaining curves correspond to the scalar-modified profiles
			with $\delta_{\infty}=0.1$, $0.3$, and $0.5$.
			The vertical line marks the transition radius
			$r_{\rm tr}=20\,{\rm kpc}$, defined by
			$r_V(r_{\rm tr})=r_{\rm tr}$.
			The scalar contribution enhances the circular velocity and
			progressively softens the radial decline relative to the
			Newtonian profile.
		}
		\label{fig:hernquist-rotation-curve}
	\end{figure}
	Figure \ref{fig:hernquist-rotation-curve} illustrates how the rotation
	curve departs from the Newtonian Hernquist behavior as the Vainshtein
	transition is approached. As the screening weakens, the scalar contribution increases the circular velocity and leads to a slower radial decline than in the Newtonian Hernquist case.
	
   For the representative parameter values considered here, however, this 
   modification does not produce an extended flat rotation curve. At larger 
   radii, the circular velocity remains decreasing, consistent with the 
   approximately $1/r$ behavior of the contributions to $v_c^2$ in the 
   outer regime. Therefore, within the regime and representative parameter 
   choice considered here, the Vainshtein-mediated scalar contribution does 
   not provide an extended flat galaxy rotation curve. Its effect is instead 
   to enhance the circular velocity and soften the Newtonian decline around 
   the Vainshtein transition. Thus, whether the scalar contribution can 
   reproduce observed rotation curves across realistic galaxy samples remains 
   an open observational question and requires further investigation over a 
   broader range of galactic systems and model parameters.

	After examining the behavior of the scalar field in the different regimes, we now consider the constraints imposed on the model parameters by galactic dynamics and Solar-System tests. The viability of the model is controlled by the parameters $\alpha$, $K_{\rm eff}$, $\Lambda$, and $m_\phi$, which must satisfy a number of consistency conditions if the scalar field is to affect galactic dynamics while remaining compatible with local gravitational constraints. A first requirement is that the transition region associated with the Vainshtein scale occurs on galactic scales, where the scalar contribution can modify the standard Keplerian behavior of the rotation curve. This requires
	\begin{equation}
		r_V(r_{\rm tr}) \sim r_{\rm tr},
	\end{equation}
	where $r_{\rm tr}$ denotes the characteristic radius of the transition region. Using the definition of the Vainshtein radius, this condition implies
	\begin{equation}
		r_{\rm tr}^3
		\sim
		\frac{8\sigma M(r_{\rm tr})}
		{\Lambda^3 K_{\rm eff}^2}.
	\end{equation}
	Thus, the combination $\frac{\sigma}{\Lambda^3 K_{\rm eff}^2}$
	is constrained by the observed galactic scale associated with the transition region. A further requirement is that the scalar field remains sufficiently light on galactic scales. Otherwise, Yukawa suppression would eliminate the scalar contribution before it can affect the rotation curve. This requires
	\begin{equation}
		m_\phi r_{\rm tr} \ll 1,
	\end{equation}
	where $r_{\rm tr}$ denotes the characteristic radius of the transition region.
	Local gravitational tests impose an additional constraint requiring the scalar-mediated force to remain strongly screened at Solar-System scales. This requires
	\begin{equation}
		r_{\rm SS} \ll r_V(M_\odot),
	\end{equation}
	where $r_{\rm SS}$ denotes a characteristic Solar System scale and $M_\odot$ denotes the solar mass.  Using the Vainshtein radius associated with the solar mass, this condition becomes
	\begin{equation}
		r_{\rm SS}^3
		\ll
		\frac{8\sigma M_\odot}
		{\Lambda^3 K_{\rm eff}^2}.
	\end{equation}
	Finally, the scalar contribution must be sufficiently large to affect galactic dynamics. This requires
	\begin{equation}
		v_c^2(r)\sim v_{\rm obs}^2(r)-v_{\rm bar}^2(r).
	\end{equation}
	The above conditions constrain the parameter combinations entering the Vainshtein scale, the scalar interaction range, and the scalar contribution to the gravitational interaction. The galactic transition scale, the screening requirement at Solar-System scales, and the observed magnitude of the rotational-velocity contribution cannot be chosen independently, but must be satisfied simultaneously within the same parameter set. Although the present analysis does not determine specific parameter values, it identifies the parameter combinations relevant for galactic dynamics and for maintaining screening at Solar-System scales. A detailed comparison with observed rotation curves and a quantitative analysis based on specific Solar-System tests are left for future work.
	
	The scalar contribution to the modified rotation curve can also be examined through the corresponding radial acceleration. In this form, the effect of the scalar field enters explicitly in the acceleration law. Such a formulation is particularly useful for examining the radial acceleration relation observed in galaxies \cite{McGaugh2016,Lelli2017}. The effective radial acceleration associated with circular motion is defined as
	\begin{equation}
		g_{\rm obs}(r) = \frac{	v_c^2(r)}{r}.
	\end{equation}
	Using the transition-region scalar-field solution in Eq.(46), the radial acceleration becomes
	\begin{equation}
		g_{\rm obs}(r)
		=
		\frac{M(r)}{G_4(0)\,r^2}
		+
		\frac{K(0)}{6G_4(0)}\,r
		-
		\frac{\sigma K_{\rm eff}\Lambda^3 r}{16}
		\left[
		-1+
		\sqrt{
			1+4\left(\frac{r_V(r)}{r}\right)^3
		}
		\right]
	\end{equation}
	It is convenient to decompose this expression as $g_{\rm obs}(r) = g_{\rm bar}(r) + g_K(r) + g_\phi(r)$, where the individual contributions are given by
	\begin{align}
		g_{\rm bar}(r) &= \frac{M(r)}{G_4(0)\,r^2}, \\
		g_K(r) &= \frac{K(0)}{6G_4(0)}\,r, \\
		g_\phi(r) &=
		-
		\frac{\sigma K_{\rm eff}\Lambda^3 r}{16}
		\left[
		-1+
		\sqrt{
			1+4\left(\frac{r_V(r)}{r}\right)^3
		}
		\right]
	\end{align}
	The sign of the scalar contribution is determined by the coupling parameter $\sigma$. For the branch considered here, the quantity
	$
	\left[
	-1+
	\sqrt{
		1+4\left(\frac{r_V(r)}{r}\right)^3
	}
	\right]
	$
	is strictly positive, so that the sign of $g_\phi$ is controlled by $-\sigma$. Therefore, for $\sigma<0$, the scalar contribution is positive and increases the total inward acceleration, leading to larger circular velocities than in the absence of the scalar interaction. From the effective potential in Eq.(57), the scalar contributionto the radial acceleration is
	\begin{equation}
		g_\phi(r) = -\frac{\sigma}{2}\,\varphi'(r).
	\end{equation}
	Therefore, the radial acceleration relation becomes
	\begin{equation}
		g_{\rm obs}(r)
		=
		g_{\rm bar}(r)
		+
		g_K(r)
		-
		\frac{\sigma}{2}\,\varphi'(r).
	\end{equation}
	The results obtained here indicate that the radial acceleration relation and the modification of galactic rotation curves arise from the same scalar-field dynamics. In both cases, the observable effects are determined by the scalar-field gradient $\varphi'(r)$. The scalar contribution can equivalently be interpreted as a modification of the effective gravitational interaction,
	\begin{equation}
		F_{\rm tot}(r)=F_{\rm GR}(r)\left[1+\Delta_{\rm H}(r)\right],
	\end{equation}
	where
	\begin{equation}
		\Delta_{\rm H}(r)
		=
		-\frac{\sigma G_4(0)K_{\rm eff}\Lambda^3 r^3}{16M(r)}
		\left[
		-1+
		\sqrt{
			1+4\left(\frac{r_V(r)}{r}\right)^3
		}
		\right]
	\end{equation}
	The quantity $\Delta_{\rm H}(r)$ plays a role analogous to the screening functions commonly used in scalar-tensor theories \cite{JainZhang2008}. In both cases, the modification remains suppressed in high-density regions and becomes significant near a characteristic transition scale \cite{Schmidt2010,Barreira2015}. In the present framework, the transition is controlled by the Vainshtein scale through the function $\Delta_{\rm H}(r)$.
	The decomposition of Eq.(81) allows the baryonic, background-curvature, and scalar contributions to the acceleration to be identified separately.
	In particular, the transition-regime solution shows that the scalar force remains finite near $r\sim r_V$, since the square-root factor remains regular at the transition scale. The relative importance of the scalar contribution can be quantified by the ratio
	\begin{equation}
		\mathcal{R}(r)
		=
		\left|
		\frac{g_\phi(r)}
		{g_{\rm bar}(r)}
		\right|.
	\end{equation}
	This quantity satisfies $\mathcal{R}(r)\ll1$ in the inner region and increases toward the transition scale, indicating the growing relative importance of the scalar contribution. At sufficiently large radii, the scalar contribution approaches the linear regime, where $g_\phi(r)\propto r^{-2}$ and consequently $v_\phi^2(r)\propto r^{-1}$. The radial dependence of $\mathcal{R}(r)$ therefore shows that the scalar contribution cannot be described as a simple rescaling of the Newtonian force, but instead evolves continuously from the screened to the linear regime through the Vainshtein transition. Consequently, the modification of the radial acceleration relation is determined by the local scalar-field configuration.
	
	\subsection{Effective Dynamical and Lensing Masses}
	In this section, we examine the effective dynamical and lensing masses
	in the gravitational-wave-compatible cubic Horndeski theory using the
	weak-field solutions derived in Sec.II. Our aim is to examine how the nonlinear scalar interaction affects the difference between the masses inferred from nonrelativistic dynamics and light propagation across the screened, transition, linear, and Yukawa regimes. In the weak-field limit, the metric in isotropic coordinates takes the form
   \begin{equation}
	ds^2=-(1+2\Phi)\,dt^2+ (1-2\Psi)\left(dr^2+r^2d\Omega^2\right).
   \end{equation}
   Following the coordinate transformations detailed in our previous works \cite{Hatice2,HaticeGW}, and using the weak-field solution for the metric perturbations obtained in Cartesian coordinates in Eq.(11), the metric can be recast in isotropic coordinates as
   \begin{equation}
	ds^2 =
	-\left(
	1-\frac{2m}{G_4(0)r}
	+\frac{K(0)}{6G_4(0)}r^{\,2}
	-\sigma\varphi
	\right)dt^{\,2}
	+
	\left(
	1+\frac{2m}{G_4(0)r}
	+\frac{K(0)}{12G_4(0)}r^{\,2}
	-\sigma\varphi
	\right)
	\left(dr^{\,2}+r^{\,2}d\Omega^2\right).
\end{equation}
	Having obtained the weak-field metric in isotropic coordinates, we now turn to the dynamics of massive particles. In particular, the circular velocity of massive particles is determined by the gravitational potential $\Phi$,
	\begin{equation}
		v_c^2(r)=r\,\Phi'(r),
	\end{equation}
	from which the effective dynamical mass is defined as
	\begin{equation}
		M_{\rm dyn}(r)=G_4(0)\,r^2\Phi'(r).
	\end{equation}
Thus, in addition to the standard Newtonian contribution, the scalar contribution to $\Phi$ modifies the mass inferred from the motion of massive particles. The effective lensing potential, on the other hand, depends on both metric potentials through the Weyl potential. Accordingly, in isotropic coordinates, the effective lensing mass inferred from the metric potentials is defined as
	\begin{equation}
		M_{\rm lens}(r)
		=
		\frac{G_4(0)\,r^2}{2}
		\left[\Phi'(r)+\Psi'(r)\right].
	\end{equation}
From the isotropic weak-field metric above, the metric potentials are identified as
\begin{align}
	\Phi(r) &=
	-\frac{M(r)}{G_4(0)\,r}
	+\frac{K(0)}{12G_4(0)}\,r^2
	-\frac{\sigma}{2}\,\varphi(r),\\
	\Psi(r) &=
	-\frac{M(r)}{G_4(0)\,r}
	-\frac{K(0)}{24G_4(0)}\,r^2
	+\frac{\sigma}{2}\,\varphi(r).
\end{align}
The scalar field contributes with opposite signs to the two metric potentials, leading in general to a nonvanishing gravitational slip, $\Phi\neq\Psi$. This distinction affects the dynamical and effective lensing masses differently, since the scalar contribution enters the potential governing massive-particle motion while it cancels explicitly from the Weyl combination governing the effective lensing mass. Combining the definitions of the effective dynamical and lensing masses, the corresponding mass difference can be written in the general form
\begin{equation}
	M_{\rm dyn}(r)-M_{\rm lens}(r)
	=
	\frac{G_4(0)\,r^2}{2}
	\left[\Phi'(r)-\Psi'(r)\right]
	=
	-\frac{\sigma G_4(0)}{2}\,
	r^2\varphi'(r)
	+
	\frac{K(0)}{8}\,r^3 .
\end{equation}
Consequently, the scalar field affects the dynamical and effective lensing masses differently, reflecting the gravitational slip between the two metric potentials \cite{JainZhang2008,Schmidt2010}. This motivates a comparison of the gravitational effects inferred from massive-particle motion and light propagation, particularly in the present case where the nonlinear scalar interaction changes its strength across the Vainshtein scale. The distinction between these two mass estimates has a well-established basis in the literature. Faber and Visser \cite{FaberVisser2006} formulated dynamical and lensing observations in terms of distinct effective, or pseudo-mass, quantities for static
spherically symmetric systems. More recent modified-gravity analyses have formulated this distinction directly in terms of the metric potentials, defining an effective lensing mass through the Weyl potential \cite{Pizzuti2019,Pizzuti2021}. The definitions in Eqs.(82)-(83) follow this metric-based prescription. In the present theory, the resulting dynamical-lensing mass difference therefore allows us to examine how the scalar modification changes between the screened, transition, and unscreened regions. Combined analyses of gravitational lensing and stellar dynamics have also been used to place observational constraints on deviations from General Relativity \cite{Collett2018}.

Unlike the matter-sourced scalar contribution, the term proportional to K(0) arises from the background curvature.  Its role in the effective lensing mass is related to the long-standing discussion of light propagation in non-asymptotically flat spacetimes. Rindler and Ishak \cite{RindlerIshak2007} showed that a cosmological term can affect the physically measured deflection
through the spacetime geometry, even when it does not explicitly appear
in the photon orbit equation. A similar treatment was adopted in our previous scalar-tensor analysis \cite{Hatice2}, where a cosmological term or a nonvanishing minimum of the scalar potential was found to contribute to light deflection. For this reason, the background-curvature contribution is retained in the general effective-mass expressions, while its relevance is determined separately in each radial regime. Its relevance, however, depends on the radial regime under consideration and on the approximations used for the corresponding scalar-field solution. In particular, it can be neglected where the local matter contribution dominates, whereas it must be retained when the background term is no longer subdominant. We therefore retain the $K(0)$ term in the effective lensing mass, consistently accounting for the influence of the background curvature on light propagation through the spacetime geometry. When the cosmological background is separated from the local lensing contribution, the homogeneous background term does not enter the local lensing potential \cite{SimpsonPeacockHeavens2010}

The mass difference acquires a different radial behavior depending on the  scalar-field solution in each regime. We therefore examine its behavior separately in the screened, transition, intermediate, and outer regions.
In the screened regime, $r\ll r_V$, the background-curvature contribution is negligible compared with the local matter source. Using the screened scalar-field solution, the dynamical-lensing mass difference becomes
\begin{equation}
	M_{\rm dyn}(r)-M_{\rm lens}(r)
	\simeq
	-\frac{\sigma G_4(0)}{2}
	\left(
	-\frac{\sigma M(r)}{\beta}
	\right)^{1/2}
	r^{3/2}.
\end{equation}
For the parameter range considered here, $\sigma<0$, the screened-regime
mass difference is positive, implying
\begin{equation}
	M_{\rm dyn}(r)>M_{\rm lens}(r).
\end{equation}
This difference, however, decreases toward the inner part of the screened region. For a fixed enclosed mass, it scales as $r^{3/2}$ and tends to zero as $r/r_V\rightarrow0$. Thus, although the dynamical mass remains larger than the lensing mass, the two estimates approach each other as the scalar contribution is increasingly suppressed by the Vainshtein mechanism. In the linear regime,
\begin{equation}
	M_{\rm dyn}(r)-M_{\rm lens}(r)
	\simeq
	\frac{\sigma^2G_4(0)M(r)}
	{K_{\rm eff}}.
\end{equation}
Unlike in the screened regime, the scalar contribution to the dynamical-lensing mass difference is no longer suppressed by the nonlinear derivative interaction. For a fixed enclosed mass, the difference approaches a finite value determined by the effective scalar coupling. In the intermediate regime, where the background-curvature contribution proportional to $K(0)$ is retained, the dynamical-lensing mass difference becomes
\begin{equation}
	M_{\rm dyn}(r)-M_{\rm lens}(r)
	\simeq
	\frac{\sigma^2G_4(0)M(r)}
	{K_{\rm eff}}
	+
	K(0)r^3
	\left[
	\frac{1}{8}
	-
	\frac{\sigma^2G_4(0)}
	{3K_{\rm eff}}
	\right].
\end{equation}
In contrast to the linear regime, the mass difference now contains an additional background-curvature contribution with an $r^3$ dependence. Thus, while the local scalar contribution is controlled by the matter source and the effective scalar coupling, the radial behavior of the mass difference also reflects the influence of the nonvanishing background term $K(0)$. In the outer regime, the dynamical-lensing mass difference becomes
\begin{equation}
	M_{\rm dyn}(r)-M_{\rm lens}(r)
	=
	\frac{\sigma^2G_4(0)M}
	{K_{\rm eff}}
	\left(1+m_\phi r\right)e^{-m_\phi r}
	+
	\frac{K(0)}{8}r^3 .
\end{equation}
The matter-sourced scalar contribution is exponentially suppressed beyond the scalar Compton wavelength, $m_\phi^{-1}$, with its radial behavior governed by the factor $(1+m_\phi r)e^{-m_\phi r}$. The background-curvature contribution, however, is not subject to Yukawa suppression and remains in the mass difference at large radii. Thus, the scalar mass determines the range of the localized scalar
modification, while the asymptotic behavior can retain a contribution
associated with $K(0)$. Using Eq.(46), and treating $r_V$ as approximately constant across the transition region, the dynamical-lensing mass difference becomes
\begin{equation}
	M_{\rm dyn}(r)-M_{\rm lens}(r)
	=
	\frac{\sigma G_4(0)K_{\rm eff}\Lambda^3}{16}
	r^3
	\left[
	1-S(r)
	\right],
\end{equation}
where, for notational convenience, we define
\begin{equation}
	S(r)
	\equiv
	\sqrt{
		1+4\left(\frac{r_V}{r}\right)^3
	}.
\end{equation}
At $r=r_V$, one has $S(r_V)=\sqrt{5}$, and hence
\begin{equation}
	M_{\rm dyn}(r_V)-M_{\rm lens}(r_V)
	=
	\frac{\sigma G_4(0)K_{\rm eff}\Lambda^3}{16}
	r_V^3
	\left(1-\sqrt{5}\right).
\end{equation}
Thus, apart from a numerical factor of order unity, the magnitude of the mass difference near the Vainshtein scale scales as
\begin{equation}
	\left|
	M_{\rm dyn}(r_V)-M_{\rm lens}(r_V)
	\right|
	\sim
	|\sigma|\,G_4(0)K_{\rm eff}\Lambda^3r_V^3 .
\end{equation}
For the attractive branch considered in this work, $\sigma<0$, and assuming
$G_4(0)>0$, $K_{\rm eff}>0$, and $\Lambda^3>0$, the above expression implies
\begin{equation}
	M_{\rm dyn}(r_V)>M_{\rm lens}(r_V).
\end{equation}
Thus, near the Vainshtein scale, the effective mass inferred from nonrelativistic dynamics exceeds that inferred from gravitational lensing. In contrast to the strong suppression found deep inside the Vainshtein region, the mass difference becomes appreciable around $r\sim r_V$ as the nonlinear screening weakens. The transition region therefore identifies the radial scale over which the scalar contribution begins to produce a significant separation between the two mass
estimates. In the limit $\beta\rightarrow0$, one has $r_V\rightarrow0$, and the
screened and transition regimes are no longer present. The dynamical-lensing mass difference then reduces to the linear-regime result given in Eq.(94), corresponding to the linear Horndeski behavior in the absence of Vainshtein screening.
	
In General Relativity, the equality of the metric potentials leads to
consistent dynamical and lensing mass estimates, $M_{\rm dyn}=M_{\rm lens}$. This relation can be modified in theories with an additional scalar degree of freedom. In covariant cubic Galileon gravity, the Vainshtein mechanism suppresses
modifications to gravity at small radii, while deviations may become relevant
on larger scales, affecting the relation between dynamical and lensing
observables \cite{Barreira2015}. Differences between the two mass estimates also arise in $f(R)$ gravity through modifications of the gravitational potentials \cite{Schmidt2010}, while beyond-Horndeski interactions can modify the corresponding potentials inside extended sources \cite{KoyamaSakstein2015}. The results obtained here exhibit a similar distinction, with the dynamical-lensing mass difference changing across the different scalar-field regimes as the efficiency of Vainshtein screening and the relative importance of the background contribution vary with radius.
	
\begin{figure}[H]
	\begin{centering}
		\includegraphics[draft=false,width=7.5cm]
		{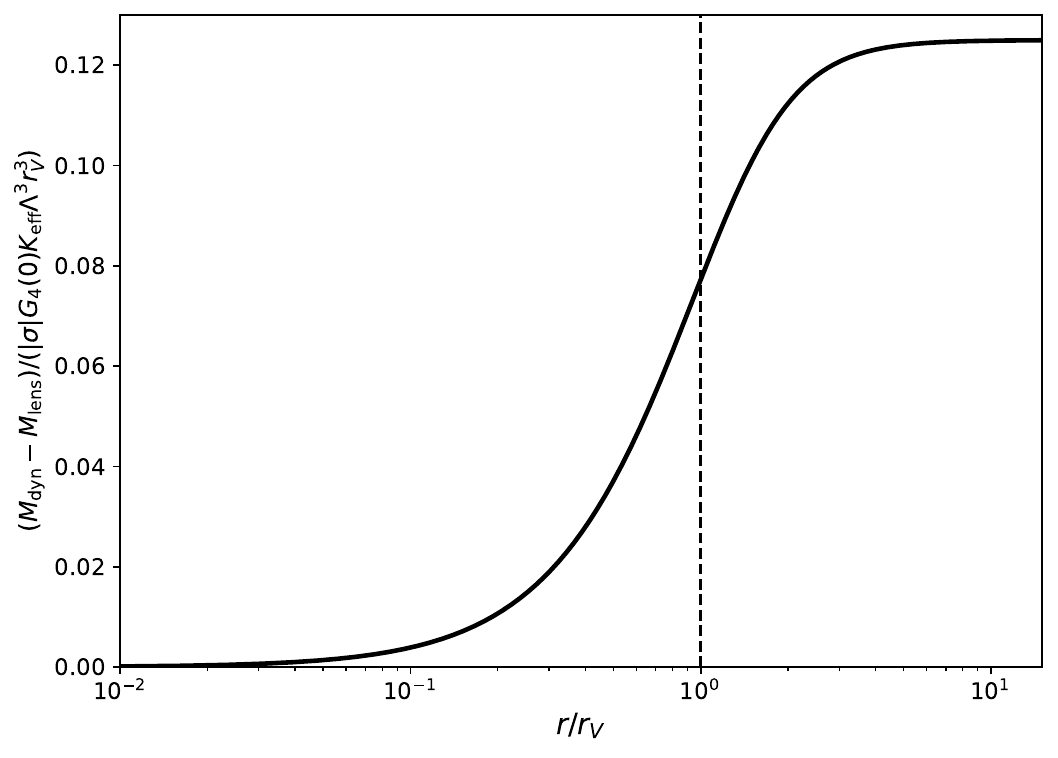}
		\caption{Dimensionless dynamical-lensing mass difference obtained
			from the transition solution as a function of the normalized radius
			$r/r_V$. The vertical dashed line indicates the Vainshtein scale.}
		\label{transition_lensing_mass}
	\end{centering}
\end{figure}

Using the transition solution derived above, we examine the radial
behavior of the dynamical-lensing mass difference around the
Vainshtein scale. Figure\ref{transition_lensing_mass} shows the
dimensionless mass difference as a function of the normalized radius
$r/r_V$.
The mass difference remains strongly suppressed in the screened region and increases as the Vainshtein scale is approached. Around the transition region, the dynamical and lensing mass estimates begin to separate, with the dynamical mass exceeding the lensing mass for the parameter branch considered here. The continuation of the transition solution beyond $r_V$ indicates the weakening of the Vainshtein suppression, although its large-radius behavior should not be interpreted as the asymptotic solution. At larger distances, the radial behavior is instead described by the corresponding intermediate and outer solutions.
Beyond the scalar Compton wavelength, the matter-sourced scalar contribution is exponentially suppressed by the Yukawa factor, while the background-curvature contribution may remain when $K(0)\neq0$.

Using the definition of the Vainshtein radius, the combination $K_{\rm eff}\Lambda^3 r_V^3$ can be expressed in terms of the scalar coupling parameters. The dynamical-lensing mass difference near the Vainshtein scale is therefore governed by the same parameters that control the scalar contribution to the gravitational interaction. These parameters also enter the modification of the rotation curve, although the two observables reflect different aspects of the scalar interaction. The rotation curve is determined by the scalar contribution
to the radial acceleration, while the dynamical-lensing mass difference reflects the different ways in which the scalar field enters the metric potentials.

\subsection{Effective Potential}
For the static and spherically symmetric metric, the radial motion of a
test particle is governed by the geodesic equation
\begin{equation}
	\ddot{r}
	+\Gamma^r_{\mu\nu}\dot{x}^{\mu}\dot{x}^{\nu}=0.
\end{equation}
In the nonrelativistic weak-field regime, where $v\ll 1$ and the metric
perturbations remain small, the radial geodesic equation reduces to
\begin{equation}
	\frac{d^2r}{dt^2}
	=
	-\frac{A'}{2B}.
\end{equation}
Substituting the weak-field metric functions in Eq.(55) gives
\begin{equation}
	\ddot r
	=
	-\frac{M}{G_4(0)\,r^2}
	-\frac{K(0)}{6G_4(0)}\,r
	+\frac{\sigma}{2}\,\varphi'(r).
\end{equation}
Using the transition-region scalar-field gradient given in Eq.(46) into Eq.(104) gives
\begin{equation}
	\ddot r_{\rm tr}
	=
	-\frac{M}{G_4(0)r^2}
	-\frac{K(0)}{6G_4(0)}r
	+
	\frac{\sigma K_{\rm eff}\Lambda^3 r}{16}
	\left[
	-1+
	\sqrt{
		1+4\left(\frac{r_V}{r}\right)^3
	}
	\right].
\end{equation}
Since a simple closed-form expression for $\varphi(r)$ is not generally available in the transition regime, the corresponding effective potential is defined directly through the radial acceleration as
\begin{equation}
	\ddot r_{\rm tr}
	=
	-\frac{dV_{\rm eff}^{\rm tr}}{dr}.
\end{equation}
It follows that
\begin{equation}
	V_{\rm eff}^{\rm tr}(r)
	=
	-\int^r \ddot{\bar r}_{\rm tr}\,d\bar r,
\end{equation}
which yields
\begin{equation}
	V_{\rm eff}^{\rm tr}(r)
	=
	-\frac{M}{G_4(0)r}
	+
	\frac{K(0)}{12G_4(0)}r^2
	-
	\frac{\sigma K_{\rm eff}\Lambda^3}{16}
	\int^r
	\bar r
	\left[
	-1+
	\sqrt{
		1+4\left(\frac{r_V}{\bar r}\right)^3
	}
	\right]
	d\bar r .
  \end{equation}
Since the transition-region solution is known directly in terms of $\varphi'(r)$, the scalar contribution to the effective potential is naturally expressed in integral form. Around the Vainshtein scale, the scalar force is neither fully screened nor purely linear, and its contribution can become comparable to the Newtonian attraction. The resulting behavior of the effective potential is therefore determined by the evolution of the scalar field gradient across the transition region. 
	
The behavior of the solution can be understood by examining the asymptotic limits on either side of the Vainshtein scale. These limits correspond to the screened $(r\ll r_V$) and linear $(r\gg r_V$) regimes on either side of the Vainshtein scale. For this purpose, the enclosed mass is assumed to vary slowly throughout the transition region, allowing $M(r)$ to be approximated by its local value and the corresponding Vainshtein radius to be treated as approximately constant. For \(r\ll r_V\), one has  
	\begin{equation}
		\sqrt{
			1+4\left(\frac{r_V}{r}\right)^3
		}
		\simeq
		2\left(\frac{r_V}{r}\right)^{3/2}.
	\end{equation}
	The scalar-field derivative therefore reduces to
	\begin{equation}
		\varphi'_{\rm tr}(r)
		\simeq
		\frac{K_{\rm eff}\Lambda^3}{4}
		r_V^{3/2} r^{-1/2},
	\end{equation}
	which implies the scaling $\varphi_{\rm tr}(r)\propto r^{1/2}$. 
	Consequently, the scalar contribution to the effective potential behaves as
	\begin{equation}
		V_{\phi}^{\rm tr}(r)\propto -\sigma \sqrt r .
	\end{equation}
	The \(\sqrt r\) scaling differs qualitatively from the Newtonian 1/r behavior. Consequently, the scalar field does not merely rescale the Newtonian potential but introduces a different radial dependence in the screened regime. In the opposite limit, $r \gg r_V$, the quantity $(r_V/r)^3$ becomes small and the square root can be expanded as
	\begin{equation}
		\sqrt{
			1+4\left(\frac{r_V}{r}\right)^3
		}
		\simeq
		1+2\left(\frac{r_V}{r}\right)^3 .
	\end{equation}
	This gives
	\begin{equation}
		\varphi'_{\rm tr}(r)
		\simeq
		\frac{K_{\rm eff}\Lambda^3}{4}
		\frac{r_V^3}{r^2}.
	\end{equation}
	Thus, on the linear side of the transition, the scalar-field gradient follows
	the same $r^{-2}$ radial scaling as the Newtonian force. Accordingly, the
	scalar contribution to the effective potential approaches a $1/r$ behavior.
	The effective-potential analysis therefore shows that the scalar contribution
	changes from a non-Newtonian $\sqrt{r}$ behavior inside the Vainshtein radius
	to an approximately Newtonian $1/r$ scaling outside it. The transition between
	these two radial behaviors occurs at the Vainshtein scale. In the limit
	$\beta\rightarrow0$, one has $r_V\rightarrow0$, so that the screened and
	transition regimes are no longer present. The scalar-field gradient then
	follows the linear behavior $\varphi'(r)\propto M(r)/r^2$, with the
	corresponding scalar contribution to the effective potential scaling as
	$M(r)/r$.
	
	The effective potential defined above describes the gravitational contribution to the radial motion. To analyze circular orbits, the centrifugal contribution associated with the conserved angular momentum L must also be included. The corresponding orbital effective potential is therefore given by
	\begin{equation}
		V_{\rm orb}^{\rm tr}(r)
		=
		V_{\rm eff}^{\rm tr}(r)
		+
		\frac{L^2}{2r^2}.
	\end{equation}
	Circular orbits occur at radii satisfying
	\begin{equation}
		\frac{dV_{\rm orb}^{\rm tr}}{dr}\Big|_{r=r_c}=0,
	\end{equation}
	while stability requires
	\begin{equation}
		\frac{d^2V_{\rm orb}^{\rm tr}}{dr^2}\Big|_{r=r_c}>0.
	\end{equation}
	Because the effective potential is given in an integral form, an exact analytic treatment of the circular-orbit conditions is generally difficult. Nevertheless, the asymptotic limits discussed above provide qualitative information about the orbital structure. In particular, for $r \ll r_V$, the scalar contribution introduces a non-Newtonian $\sqrt r$ correction, which modifies the shape of the effective potential relative to the Newtonian case and may shift the radii of circular orbits. In the asymptotic limit \(r\ll r_V\), the orbital effective potential takes the approximate form
	\begin{equation}
		V_{\rm orb}^{\rm tr}(r)
		\sim
		-\frac{M}{G_4(0)r}
		+
		A\sqrt r
		+
		\gamma r^2
		+
		\frac{L^2}{2r^2},
	\end{equation}
	where
	\begin{equation}
		A
		\equiv
		-\frac{\sigma K_{\rm eff}\Lambda^3}{4}
		r_V^{3/2},
		\qquad
		\gamma
		\equiv
		\frac{K(0)}{12G_4(0)} .
	\end{equation}
	The term \(A\sqrt r\) represents the scalar contribution arising from the screened asymptotic behavior of the transition solution. For the attractive branch with \(\sigma<0\), one has \(A>0\), so this term gives an additional attractive radial force proportional to \(r^{-1/2}\). This radial dependence is weaker than the Newtonian \(r^{-2}\) force and therefore modifies the shape of the effective potential relative to the purely Newtonian case. The term \(\gamma r^2\) is associated with the background-curvature contribution proportional to \(K(0)\) and becomes relevant only at sufficiently large distances. The resulting orbital structure is therefore determined by the combined effect of Newtonian attraction, the scalar correction, the background-curvature term, and the centrifugal barrier. Using the approximate asymptotic form of the effective potential, the circular-orbit condition becomes
	\begin{equation}
		\frac{M}{G_4(0)r_c^2}
		+\frac{A}{2\sqrt{r_c}}
		+2\gamma r_c
		-\frac{L^2}{r_c^3}
		=0,
	\end{equation}
	which gives
	\begin{equation}
		L^2
		=
		\frac{M r_c}{G_4(0)}
		+\frac{A}{2}r_c^{5/2}
		+2\gamma r_c^4.
	\end{equation}
	The stability condition in Eq.(116) takes the form
	\begin{equation}
		-\frac{2M}{G_4(0)r_c^3}
		-\frac{A}{4r_c^{3/2}}
		+2\gamma
		+\frac{3L^2}{r_c^4}
		>0.
	\end{equation}
	These expressions show that the scalar contribution modifies both the circular-orbit condition and the stability criterion relative to the standard Newtonian case. In particular, the term proportional to \(A\) introduces a non-Newtonian correction associated with the screened-side behavior of the transition solution. The corresponding scalar force scales as \(r^{-1/2}\), whereas the Newtonian force scales as \(r^{-2}\). Therefore, the scalar force decreases more slowly with radius than the Newtonian attraction and can change the balance between gravitational attraction and centrifugal support near the transition region.
	The circular-orbit condition shows that the scalar contribution increases the angular momentum required to sustain circular motion at a given radius. The stability condition also acquires an additional term proportional to $A/r_c^{3/2}$, indicating that the scalar field modifies the local curvature of the effective potential. Since this contribution enters with a negative sign, it acts in the direction of reducing the curvature associated with stable circular motion. The contribution proportional to \(\gamma\) becomes important only at sufficiently large radii, where the background-curvature term associated with K(0) becomes comparable to the remaining contributions. Near the Vainshtein scale, the orbital behavior is determined primarily by the Newtonian and scalar contributions. These results show that the transition-region scalar field influences not only the rotation curve but also the structure of circular orbits through its contribution to the effective potential.

	\section{Inter-Galactic dynamics}
	In this section, we investigate intergalactic dynamics using the outer-region weak-field solution derived in Sec. II, which is appropriate for distances beyond the galactic scale. In particular, we analyze the relative radial acceleration and derive the corresponding turnaround radius in both the Yukawa-suppressed and long-range regimes. This allows us to determine how the scalar interaction modifies gravitational dynamics on intergalactic scales.
	\subsection{Relative Acceleration Parameter}
	We first derive the relative radial acceleration using the outer-region weak-field solution, which is appropriate for intergalactic distances. In the intergalactic regime, the separation between galaxies is much larger than their characteristic sizes, so that the details of their internal mass distributions can be neglected at the scales of interest. Each galaxy can therefore be treated, to leading order, as an effectively point-like gravitating source characterized by its total mass. In the regime $r \gg r_V$, the nonlinear derivative self-interactions proportional to $\beta$ become subdominant, and the scalar field is described by the linearized outer-region solution given in Eq.(51). At distances exceeding the scalar Compton wavelength, $r \gg m_\phi^{-1}$, the Yukawa contribution associated with the localized source becomes exponentially suppressed, and the scalar field approaches the constant value determined by K(0). The intergalactic dynamics in this regime can therefore be examined through the corresponding large-distance form of the weak-field metric. In the weak-field and nonrelativistic limit, substituting the large-distance scalar-field solution given in Eq.(51) into the radial equation of motion derived in Eq.(104), and retaining the leading weak-field contributions, yields the relative radial acceleration of the two galaxies as
   \begin{equation}
	\frac{d^2r}{dt^2}
	=
	-\frac{M}{G_4(0)\,r^2}
	\left[
	1+\eta(1+m_\phi r)e^{-m_\phi r}
	\right]
	-\frac{K(0)}{6G_4(0)}\,r ,
   \end{equation}
   Here, r denotes the radial separation between the galaxies and M denotes their total mass. For convenience, we define the dimensionless coupling parameter
	\begin{equation}
		\eta
		\equiv
		\frac{\sigma^2 G_4(0)}
		{K_{\rm eff}}=
		\frac{G_{4,\phi}^2(0)}
		{G_4(0)K_{,X}(0)+3G_{4,\phi}^2(0)}
	\end{equation}
	The Yukawa contribution increases the effective attractive interaction at distances smaller than the scalar Compton wavelength, while becoming exponentially suppressed for $m_\phi r\gg1$. At sufficiently large distances, the dynamics are therefore governed by the competition between the Newtonian attraction and the background-curvature contribution associated with $K(0)$. Since $G_{4,\phi}^2(0)\ge0$, and physically viable models require $G_4(0)>0$ and $K_{,X}(0)>0$, the parameter $\eta$ satisfies
	\begin{equation}
		0\le \eta < \frac{1}{3}.
	\end{equation}
	Therefore, $\eta$ can be interpreted as a bounded dimensionless parameter that quantifies the relative strength of the scalar contribution to the effective gravitational interaction. While $\eta$ is constrained by the theoretical consistency conditions of the theory, the remaining parameters determine the amplitude and spatial range of the scalar-field effects. Parameters of this type frequently appear in studies of screened fifth forces and astrophysical tests of modified gravity \cite{Jain2013,Sakstein2015,Desmond2018,Burrage2018}.
	
	The behavior of the system is governed by the ratio between the radial distance \(r\) and the scalar Compton wavelength \(m_\phi^{-1}\). Two distinct regimes can therefore be identified. For \(m_\phi r\ll1\), the scalar interaction remains effective and the system is in the long-range regime. For $m_\phi r\gg1$, the scalar contribution becomes exponentially suppressed, corresponding to the Yukawa-suppressed regime. As shown below, the parameter $\eta$ determines the relative strength of the scalar correction in the long-range limit, while the scalar mass \(m_\phi\) controls the distance scale over which this correction remains significant. These two limits lead to qualitatively different gravitational behaviors and provide a useful framework for analyzing the large-scale dynamics predicted by the theory. In the long-range regime, defined by $m_\phi r\ll1$, the Yukawa factor satisfies
	\[
	(1+m_\phi r)e^{-m_\phi r}\simeq1 .
	\]
	The relative radial acceleration given by Eq.(113) therefore reduces to
	\begin{equation}
		\frac{d^2r}{dt^2}
		=
		-\frac{M}{G_4(0)r^2}(1+\eta)
		-
		\frac{K(0)}{6G_4(0)}\,r .
	\end{equation}
	The relative importance of the background-curvature contribution with respect to the attractive long-range gravitational interaction can be quantified by the dimensionless ratio
	\begin{equation}
		q_{\rm LR}(r)
		\equiv
		\frac{|a_K|}{|a_{\rm grav}|}
		=
		\frac{|K(0)|}
		{6M(1+\eta)}
		r^3 .
	\end{equation}
	The regime \(q_{\rm LR}\ll1\) corresponds to dynamics dominated by the attractive gravitational interaction, whereas for \(q_{\rm LR}\gtrsim1\) the background-curvature contribution becomes dynamically significant.
	The relative acceleration parameter provides a dimensionless measure of the importance of the background-curvature contribution relative to the effective gravitational attraction. Consequently, it allows the transition between the attraction-dominated and background-dominated regimes to be identified directly from its radial behavior. In the long-range regime, the Yukawa factor approaches unity and the scalar contribution is no longer exponentially suppressed. The scalar force therefore follows the same $r^{-2}$ radial dependence as the Newtonian interaction. As a result, the scalar field does not introduce a new radial dependence in the force law, but instead rescales the overall strength of the effective gravitational attraction through the parameter $\eta$. Furthermore, the relative acceleration parameter exhibits a scale-dependent behavior. Since $q_{\rm LR}(r)\propto r^3$, the background-curvature contribution associated with $K(0)$ becomes increasingly important with increasing distance and eventually dominates over the effective gravitational attraction. Consequently, the large-scale dynamics is ultimately governed by the background-curvature term.
	
	In the Yukawa-suppressed regime, defined by $m_\phi r\gg1$, the exponential factor becomes negligible and the scalar contribution is exponentially suppressed. Consequently, the Yukawa correction no longer contributes appreciably to the gravitational potential, leaving only the Newtonian mass-dependent term in the weak-field metric. The normalization of $G_4(0)$ is then fixed by requiring that the $g_{00}$ component reproduces the standard Newtonian potential, yielding $G_4(0)=1$. The dynamics is therefore governed predominantly by the Newtonian attraction and the background-curvature contribution associated with $K(0)$. The radial acceleration reduces to
	\begin{equation}
		\frac{d^2 r}{dt^2}
		\simeq
		-\frac{M}{r^2}
		-
		\frac{K(0)}{6}r .
	\end{equation}
	The relative importance of the background-curvature contribution with respect to the Newtonian attraction can be characterized by the dimensionless ratio
	\begin{equation}
		q_{\rm YR}(r)
		\equiv
		\frac{|a_K|}{|a_N|}
		=
		\frac{|K(0)|}{6M}r^3 .
	\end{equation}
	Since $q_{\rm YR}(r)\propto r^3$, the relative importance of background-curvature term increases with distance. Consequently, at sufficiently large intergalactic scales, the background-curvature contribution eventually dominates the dynamics. 
	
    A similar limiting behavior was found in our previous analysis of Brans-Dicke
    theory with a scalar potential \cite{Hatice2}. For a light scalar, the Yukawa interaction remains long ranged and modifies the coefficient of the $r^{-2}$ contribution, whereas for a heavy scalar the exponential correction is suppressed. Thus, in both cases, the scalar interaction exhibits a long-range regime in which the scalar field modifies the gravitational force and a short-range regime in which the scalar contribution is exponentially suppressed, leaving only the Newtonian and background contributions. The two results differ in the strength of the scalar correction and in the
    form of the scalar-field solution. In the Brans-Dicke case \cite{Hatice2},
    the Yukawa correction is controlled by the parameter $\omega$, whereas in
    the present Horndeski model its strength is determined by $\eta$. The former
    is based on a point-mass scalar-field solution, while the present result is
    obtained from the outer scalar-field solution. When the Yukawa contribution is suppressed, the remaining inverse-square  term recovers the standard Newtonian radial dependence, while the term  proportional to $r$ represents the additional background contribution retained in the present weak-field description \cite{Faraoni2015}.
    
	\subsection{Turnaround radius and binding condition}
	Using the modified radial acceleration derived in the previous subsection, we determine the turnaround radius of gravitationally bound systems. Since the turnaround scale occurs at intergalactic distances, where the weak-field approximation remains valid, the turnaround condition is obtained by imposing
	\begin{equation}
		\frac{d^2r}{dt^2}=0,
	\end{equation}
	which corresponds to the balance between the attractive gravitational interaction and the repulsive background-curvature contribution. It therefore determines a characteristic scale beyond which gravitationally bound structures can no longer remain stable against the background expansion. The turnaround radius has been extensively studied in the context of large-scale structure formation and modified gravity theories \cite{Pavlidou2014,Tanoglidis2015,Pavlidou2018, TurnaroundHorndeski}. Since the scalar interaction exhibits qualitatively different behaviors in the long-range and Yukawa-suppressed regimes, the corresponding turnaround scale is expected to depend on the relative importance of the scalar contribution. It is therefore useful to examine the turnaround condition separately in these two regimes in order to determine how the scalar field influences the maximum scale of gravitationally bound structures.
	In the long-range regime, defined by $m_\phi r\ll1$, the scalar contribution remains unsuppressed and modifies the effective Newtonian attraction through the factor $(1+\eta)$. The radial acceleration is therefore given by
	\begin{equation}
		\frac{d^2r}{dt^2}
		=
		-\frac{M}{r^2}(1+\eta)
		-
		\frac{K(0)}{6}r .
	\end{equation}
	Imposing the turnaround condition $d^2r/dt^2=0$ gives
	\begin{equation}
		r_t^{\rm LR}
		=
		\left[
		\frac{6M(1+\eta)}
		{|K(0)|}
		\right]^{1/3},
	\end{equation}
	where $K(0)<0$ has been assumed so that the background-curvature contribution acts repulsively and a finite turnaround radius exists. The turnaround radius in the long-range regime depends explicitly on the scalar parameter $\eta$. The scalar field therefore modifies the scale at which the attractive gravitational interaction is balanced by the background-curvature contribution. In the opposite limit, $m_\phi r\gg1$, the scalar interaction becomes exponentially suppressed and its contribution to the radial acceleration becomes negligible. In this Yukawa-suppressed regime, the mass-dependent term in the weak-field metric reduces to $2M/[G_4(0)r]$. Requiring this term to reproduce the standard Newtonian potential fixes the normalization $G_4(0)=1$ in the units adopted here. The turnaround scale is then determined solely by the Newtonian attraction and the background-curvature contribution, yielding
	\begin{equation}
		r_t^{\rm YR}
		=
		\left[
		\frac{6M}
		{|K(0)|}
		\right]^{1/3}.
	\end{equation}
	In the GR limit, the background quantity $K(0)$ is related to the cosmological constant through $K(0)\rightarrow-2\Lambda_{\rm cos}G_4(0)$. Using the Newtonian
	normalization $G_4(0)=1$, the Yukawa-suppressed turnaround radius
	reduces to
	\begin{equation}
		r_t^{\rm GR}
		=
		\left(
		\frac{3M}
		{\Lambda_{\rm cos}}
		\right)^{1/3}.
	\end{equation}
	This expression represents the GR limiting case of the general turnaround-radius relation. Since the observed value of the cosmological constant is known, the turnaround radius is completely determined by the enclosed mass M. The above expression therefore provides a direct quantitative prediction that can be compared with observational determinations of turnaround scales. Restoring physical units, the turnaround radius in the GR limit is
	\begin{equation}
		r_t
		=
		\left(
		\frac{3G_NM}
		{\Lambda_{\rm cos}c^2}
		\right)^{1/3},
	\end{equation}
	where $G_N$ is Newton's gravitational constant, $c$ is the speed of light, and $\Lambda_{\rm cos}=1.1056\times10^{-52}\,\mathrm{m}^{-2}$ is the observed cosmological constant. For a representative galactic mass,
	\begin{equation}
		M=10^{12}M_\odot
		=
		1.98847\times10^{42}\,\mathrm{kg},
	\end{equation}
	one obtains
	\begin{equation}
		r_t
		\simeq
		1.11\,\mathrm{Mpc}.
	\end{equation}
	Since the turnaround radius scales as $r_t\propto M^{1/3}$, representative values for typical astrophysical systems are listed in Table\ref{tab:turnaround_predictions}.
	\begin{table}[H]
		\centering
		\caption{Representative turnaround radii in the GR limit,
			computed using the observed cosmological constant
			$\Lambda_{\rm cosm}=1.1056\times10^{-52}\,\mathrm{m}^{-2}$.}
		\label{tab:turnaround_predictions}
		\begin{tabular}{cc}
			\hline\hline
			Enclosed mass $M$ & $r_t$ (Mpc) \\
			\hline
			$10^{12}M_\odot$        & 1.11 \\
			$2\times10^{12}M_\odot$ & 1.40 \\
			$10^{13}M_\odot$        & 2.39 \\
			$10^{15}M_\odot$        & 11.09 \\
			\hline\hline
		\end{tabular}
	\end{table}
	The numerical estimates presented in Table \ref{tab:turnaround_predictions}
	are valid only in the Yukawa-suppressed regime ($m_\phi r\gg1$), where the
	scalar interaction is exponentially suppressed and the GR limit is
	recovered. They therefore represent predictions for this limiting case,
	rather than for the full theory.
	
	In this context, these values provide theoretical predictions that can be directly compared with observational determinations of turnaround radii. Within the $\Lambda$CDM framework, the maximum turnaround radius has been proposed as a direct local test of cosmology, since it depends only on the enclosed mass and the cosmological constant \cite{Pavlidou2014,Faraoni2016}. The numerical estimates obtained in the Yukawa-suppressed regime are consistent with the standard $\Lambda$CDM turnaround relation and the typical scales reported for astrophysical systems of comparable masses. Furthermore, numerical simulations have confirmed the characteristic scaling $r_t\propto M^{1/3}$ for halos with masses $M\gtrsim10^{13}M_\odot$, while galaxy groups of this mass scale have been identified as particularly promising targets for observational tests of the turnaround bound \cite{Korkidis2020,Tanoglidis2015}.
	
	On the other hand, the turnaround radii obtained in the two limiting regimes satisfy the relation
	\begin{equation}
		r_t^{\rm LR}
		=
		r_t^{\rm YR}
		(1+\eta)^{1/3}.
	\end{equation}
	This relation shows explicitly how the scalar interaction modifies the turnaround scale relative to the Yukawa-suppressed regime. Since physically viable models satisfy \(0\le\eta<1/3\), the ratio \(r_t^{\rm LR}/r_t^{\rm YR}\) lies between \(1\) and \((4/3)^{1/3}\simeq1.10\). Thus, the scalar field increases the turnaround radius in the long-range regime, although the effect remains moderate throughout the allowed parameter range. Even for \(\eta\rightarrow1/3\), the turnaround radius exceeds the Yukawa-regime value by at most about \(10\%\). For weak scalar corrections, $\eta\ll1$, the turnaround radius can be expanded as
	\begin{equation}
		r_t^{\rm LR}
		\simeq
		r_t^{\rm YR}
		\left(
		1+\frac{\eta}{3}
		\right).
	\end{equation}
	This expression shows that the scalar field increases the turnaround radius relative to the Yukawa-suppressed regime. For $\eta\ll1$, the effect remains perturbatively small, with a fractional correction of approximately $\eta/3$. The dependence on $\eta$ arises from the scalar contribution to the attractive interaction in the long-range regime, which increases the scale at which the attractive and background-curvature contributions become comparable. The same parameter that controls the scalar contribution to the rotation curves and the lensing-dynamical mass difference also enters the turnaround radius, indicating that these observables are influenced by the same underlying scalar interaction. In the Yukawa-suppressed regime, by contrast, the scalar contribution becomes exponentially negligible and the turnaround scale approaches the value determined solely by the Newtonian attraction and the background-curvature contribution. These results show that the effects of the scalar field are not restricted to galactic dynamics, but can also modify the characteristic scales associated with gravitationally bound systems.
	
	\section{Conclusions}
	In this work, we investigated the weak-field dynamics of a gravitational-wave compatible subclass of Horndeski gravity containing nonlinear scalar interactions. The metric and scalar perturbations were assumed to remain within the weak-field regime, while the nonlinear derivative self-interactions of the scalar field were retained in order to describe the scale-dependent screening behavior associated with the Vainshtein mechanism. Within this framework, static and spherically symmetric weak-field solutions were obtained together with the corresponding scalar-field configurations and gravitational potentials across different radial regimes. The resulting dynamics exhibits a scale-dependent gravitational behavior connecting the galactic transition regime with the asymptotic intergalactic dynamics.
	
	We then investigated galactic dynamics through the motion of timelike test particles and derived the corresponding circular velocity. The resulting velocity consists of the Newtonian contribution, a curvature term associated with $K(0)$, and an additional contribution generated by the nonlinear scalar interaction. The scalar contribution becomes most significant in the vicinity of the Vainshtein scale, where it modifies the radial dependence of the gravitational force and produces departures from the standard Keplerian behavior. Outside this transition region, the scalar contribution is either screened at small radii or becomes Yukawa-suppressed at large distances, so that the scalar-induced modifications become progressively less important. The analysis therefore shows that nonlinear scalar interactions can enhance the circular velocity over a finite radial interval around the Vainshtein scale, while an approximately Newtonian radial dependence is recovered away from the transition regime. This behavior arises from the nonlinear scalar dynamics and the associated Vainshtein transition. It should be emphasized, however, that the present solution characterizes the intermediate regime and does not by itself produce the asymptotically flat rotation curves observed at large galactocentric distances.
	
	The scalar interaction influences not only the rotational dynamics of galaxies but also the relation between the motion of massive particles and the propagation of light. Owing to the gravitational slip generated by the scalar field, the dynamical mass inferred from the motion of stars and gas generally differs from the lensing mass inferred from gravitational lensing. In the present gravitational-wave compatible cubic Horndeski model, this discrepancy reaches its largest value near the Vainshtein scale, where the scalar contribution becomes comparable to the Newtonian interaction. Consequently, the transition regime simultaneously enhances the circular velocity and produces a localized lensing-dynamical mass difference, reflecting the underlying nonlinear scalar dynamics.
	
	Beyond galactic dynamics, the scalar interaction also modifies the turnaround radius of gravitationally bound systems. The resulting dependence on the scalar-field parameters distinguishes the present prediction from those of General Relativity and several alternative modified-gravity theories, extending the effects of the nonlinear scalar interaction to intergalactic scales. A quantitative comparison of this prediction with observed turnaround radii is left for future work.
	
	Overall, the results obtained in this work demonstrate how the scalar-field parameters influence gravitational phenomena from galactic to intergalactic scales. In particular, the nonlinear scalar interaction modifies the circular velocity in the transition regime, generates a localized lensing-dynamical mass difference through gravitational slip, and alters the turnaround radius of gravitationally bound systems. Together, these results illustrate how the same scalar interaction can affect gravitational phenomena across a wide range of astrophysical scales. The observational competitiveness of the present model relative to alternative modified-gravity scenarios, such as MOND, HMG, and STVG/MOG, has not been quantitatively assessed in this work. A systematic comparison using galaxy rotation curves, radial-acceleration relations, lensing constraints, and	turnaround-radius measurements therefore remains an important direction for future work and would provide a direct test of the phenomenological viability of the model.
	\section*{Supplementary information}
	There is no supplementary file regarding this paper.
	\section*{Data availability Statement}
	No data associated in the manuscript.	

\end{document}